\documentstyle[aps,pre,multicol,epsf]{revtex}
     
\newcommand \be{\begin{equation}}
\newcommand \ee{\end{equation}}
\newcommand \ba{\begin{eqnarray}}
\newcommand \ea{\end{eqnarray}}
\def\nn{\nonumber}

\begin{document}

\draft

\title{Transition from Townsend to glow discharge:\\
subcritical, mixed or supercritical characteristics}

\author{Danijela D.\ \v Sija\v ci\'c$^1$ and Ute Ebert$^{1,2}$}
\address{$^1$Centrum voor Wiskunde en Informatica,
P.O.Box 94079, 1090 GB Amsterdam, The Netherlands\\
$^2$Dept.\ Physics, Eindhoven Univ.\ of Techn., The Netherlands}

\date{\today}
\maketitle

\begin{abstract}
The transition from Townsend to glow discharge is investigated 
numerically in one space dimension in full parameter space within 
the classical model: 
with electrons and positive ions drifting in the local electric field, 
impact ionization by electrons ($\alpha$ process), secondary electron 
emission from the cathode ($\gamma$ process) and space charge effects.
We also perform a systematic analytical small current expansion about 
the Townsend limit up to third order in the current that fits our 
numerical data very well.
Depending on the two determining parameters $\gamma$ and system size $pd$, 
the transition from Townsend to glow discharge can show the textbook 
subcritical behavior, but for smaller values of $pd$, we also find 
supercritical or some unexpected intermediate ``mixed'' behavior.
Our work shows the same qualitative dependence of $U=U(I,pd)$ for fixed 
$\gamma$ as the old experiments by Pokrovskaya-Soboleva and Klyarfeld.
Furthermore, the analysis lays the basis for understanding the complex 
spatio-temporal patterns in short planar barrier discharge systems.
\end{abstract}


\begin{multicols}{2}

\section{Introduction}

Space charge effects in many cases are the first nonlinear effects
in gas discharges with increasing current. They are known
to induce the avalanche to streamer transition in transient
discharges as well as the transition from Townsend to normal
and further to abnormal glow in stationary discharges. 
Generically, nonlinear couplings in 
non-equilibrium systems lead to the formation of spontaneous 
spatio-temporal patterns. The current constriction in the normal
glow discharge as well as the longitudinal 
striations of a long positive column of a glow discharge 
\cite{Jonas1,Jonas2,Bruhn} fall into this class of phenomena.

Recently, the amazing variety of spatio-temporal patterns formed 
mainly in the transversal direction of a short dc driven 
system consisting of a gas discharge layer and a semiconductor 
layer sandwiched between two coplanar electrodes has drawn 
considerable attention 
\cite{Muenster1,Muenster2,StripeM,Muenster3,HexM,Astrov,Zigstr,Str,filStr}.
These patterns are due to the nonlinear gas discharge being
coupled to the linearly responding semiconductor. 
In particular, a negative differential conductivity of the gas 
discharge in some region of the current-voltage-characteristics is expected 
\cite{Engel,Radehaus89,Radehaus90,Phelps93I,Phelps93II,Phelps93III,Petro97} 
to play a significant role in the spontaneous formation of patterns, 
quite like in nonlinear semiconductor devices \cite{Schoell}.
Due to its geometry, modeling the system \cite{Zigstr,Str,filStr} 
as one-dimensional is a very good approximation, as long as this 
symmetry is not spontaneously broken by the intrinsic dynamics.
So as a first step of any investigation, the behavior and the resulting
current-voltage-characteristics of the purely one-dimensional 
gas discharge system have to be understood.

An investigation of the system \cite{Str} along the lines
of the textbook \cite{Raizer} shows that the pattern formation 
occurs at the space charge driven transition from Townsend to 
glow discharge. The gas dicharge layer is rather short, more precisely,
the product $pd$ of gas pressure $p$ times electrode distance $d$
is small. This raises the question of the Townsend to glow transition
for small $pd$. However, despite a history of more than 70 years, 
we are not aware of any thorough and complete study of this classical 
problem. 
Therefore, our aim in the present paper is to develop a consistent 
picture of the Townsend to glow transition in one dimension from 
analytical and numerical investigations, in particular, for short systems.


Many authors focus on quite long discharges that have a clearly
pronounced subcritical characteristics, i.e., for fixed large $pd$ and 
growing total current $I$, the voltage first decreases from the Townsend 
limit towards the normal glow regime, then it increases again in
the abnormal glow regime until heating effects become important
and the voltage again decreases towards the arc discharge.
We will not consider this last thermally driven transition at high 
currents. The initial decrease of voltage from Townsend discharge 
towards normal glow creates a regime of negative differential
conductivity, and some authors \cite{Petro97} believe that 
negative differential conductivity is generic for this system.

However, already in the early 1940'ies, e.g., in the extensive review 
by Druyvesteyn and Penning \cite{Druyve}, it was suggested that
this subcritical behavior might not be the only possible one,
but that also a monotonic increase of voltage with current was 
possible. Such a behavior we will call supercritical, in line
with modern bifurcation theory. There are early experimental
papers by Pokrovskaya-Soboleva and Klyarfeld \cite{Klyar57}
and McClure \cite{McClure} that clearly indicate a supercritical
transition for small values of $pd$ in hydrogen and deuterium
in combination with metal electrodes. Later data by the same 
authors \cite{Klyar66} is reproduced in Raizer's textbook
\cite{Raizer}, however, only for rather long systems with 
subcritical characteristics.

Theoretical insight into the question of bifurcation behavior
can be gained by analytical or numerical investigation 
of the appropriate model. The classical model contains the drift 
of charged particles in the local field, the $\alpha$-process of impact 
ionization in the bulk of the gas, the $\gamma$-process of 
secondary electron emission from the cathode, and space charge effects.

Numerical calculations date back 
to the 50'ies \cite{Crowe}, the first numerical evaluations 
using an ``electronic computer'' can be found in the early 60'ies 
in \cite{Ward61,Ward62}. In particular, in the work of Ward
\cite{Ward62}, current-voltage characteristics with or without
a region of negative differential conductivity can be found 
for different values of $pd$. However, computing power at the time
was quite restricted and hence only a few current-voltage-characteristics
were calculated. The work does not seem to have been extended significantly
lateron. We will take up the issue in Section IV.

Analytical efforts were constrained to small current expansions
about the Townsend limit.
The old German textbook of Engel and Steenbeck \cite{Engel}
contains an elegant argument that the initial increase or decrease
of the characteristics from the Townsend limit depends on
the sign of $\alpha''(E_T)$ where $\alpha(E)$ is the effective 
impact ionization coefficient as a function of the electric field $E$,
and ${}''$ denotes the second derivative evaluated at Townsend's 
breakdown field $E_T$. We recall this argument in Section III.B.
The book \cite{Engel} also gives an explicit expression for
the coefficient $c_2\propto\alpha''(E_T)$ in the expansion 
$U(I)=U_T+c_2I^2$, however, without derivation or reference. 
Exactly the same statements can be found more than 60 years later 
in Raizer's much read textbook \cite{Raizer}. 
Kolobov and Fiala
\cite{Kolobov} assume that $\alpha''=0$ marks the point where
negative differential conductivity disappears.
A similar small current expansion of the voltage about the Townsend
limit has recently been performed in \cite{Petro97}, but with 
a different result --- here the leading correction is found 
to be linear in the current rather than quadratic. None of the
two results has been compared to numerical solutions. In the present paper,
we will present yet another result for the small current expansion
and evaluate it to higher orders. Our derivation is a systematic expansion
and in very good agreement with our numerical results.

In general, our aim in the present paper is a consistent theoretical 
investigation of the simple classical model of these discharges 
treated by so many authors 
\cite{Engel,Phelps93III,Petro97,Raizer,Druyve,Crowe,Ward61,Ward62,Kolobov,Meek/Craggs}. 
The exploration of the full parameter space is possible, because 
the current-voltage characteristics in appropriate dimensionless units 
depends essentially only on two parameters: the secondary emission 
coefficent $\gamma$ and the dimensionless system size $L\propto pd$.

Of course, various extensions of the model can be considered:
particle diffusion, attachement, nonlinear particle mobilities,
a field-dependent secondary emission rate or nonlocal ionization rates.
However, e.g., Boeuf \cite{Boeuf} has argued 
that for the transition from normal to abnormal glow,
nonlocal terms in the impact ionization reaction should be
taken into account through hybrid numerical models \cite{FialaPB}, 
while in the subnormal regime between Townsend and normal glow, 
a local fluid model is considered sufficient \cite{Kolobov}. 
This supports the strategy to first seek a full understanding 
of the predictions of the classical model as a corner stone 
and starting point for any further work.

In the present work, we perform a systematic analytical expansion 
of the voltage about the Townsend limit
up to $O(I^3)$, recovering the qualitative features of the
solution from \cite{Engel,Raizer}: in particular, we find that a linear
term in current $I$ indeed is missing, and that the coefficient $c_2$ indeed
is proportional to $\alpha''(E_T)$, but with a different proportionality
constant. In fact, our coefficient $c_2$ depends strongly
on the secondary emission coefficient $\gamma$ --- it varies by almost
three orders of magnitude for $\gamma$ between $10^{-6}$ to $10^{-1}$ ---, 
while the expression given in \cite{Engel,Raizer} does not depend on
$\gamma$ at all. We also evaluate the next order $O(I^3)$.  
Our analytical result fits our numerical solutions very well within 
its range of validity. The stationary states of the pattern forming
system \cite{Str} are within the range of validity of this expansion.

Furthermore, we explore the current-voltage characteristics 
numerically beyond the range of the small current expansion in
the full parameter space.
We show that within the classical model, there is not only
the familiar subcritical bifurcation from Townsend to glow
discharge for large values of $pd$, but for sufficiently 
small values of $pd$, the bifurcation is supercritical, in agreement
with the scenario suggested by Druyvesteyn and Penning \cite{Druyve}.
Furthermore, for intermediate values of $pd$, there always 
exist completely unexpected ``mixed'' bifurcations. This surprising
finding implies that the negative differential conductivity
does not vanish when $\alpha''(E_T)=0$ in the Townsend limit, 
as most authors assume \cite{Raizer,Kolobov},
but only for smaller values of $pd$. These statements are true
for all relevant values of secondary emission $\gamma$.
Our three-dimensional plots of the voltage as a function
of dimensionless system size $L\propto pd$ and current $I$
for a given gas-electrode combination 
are done in the same manner as the old experimental plots by 
Pokrovskaya-Soboleva and Klyarfeld \cite{Klyar66}. 

The paper is organized as follows: in Section II,
we recall the classical model and its parameters, perform
dimensional analysis, and reformulate the stationary 
one-dimensional problem as a boundary condition problem. 
In Section III, we recall the Townsend limit and the classical
argument of Engel and Steenbeck on the qualitative dependence of 
the small current expansion on $\alpha''$. We then perform a new 
systematic small current expansion up to third order in the total 
current $I^3$ and determine the coefficients of the expansion explicitly.
Section IV begins with our numerical strategy and a discussion
of the parameters with their ranges. The parameter dependence
of the current-voltage characteristics on system size $L\propto pd$ 
and secondary emission coefficient $\gamma$ is first presented 
in the form of $(I,U,pd)$-plots for fixed $\gamma$ as in \cite{Klyar66}.
We then present spatial plots of electron current and field, and compare 
our numerical results to our analytical small current expansion. Finally,
we classify the bifurcation structure in the complete relevant
parameter space. Section V contains a summary and an outlook
onto the implications of this work for spatio-temporal pattern 
formation in barrier discharges. Two appendices contain the proof 
of the uniqueness of the solution of the boundary value problem 
and details of the small current expansion in order $I^3$.

\section{The classical model}

\subsection{Definition}

We investigate the classical model for glow discharges in simple 
non-attaching gases in a planar, quasi-one-dimensional geometry.
The same model was previously investigated in, e.g.,
\cite{Engel,Raizer,Druyve,Crowe,Ward61,Ward62,Kolobov,Meek/Craggs} 
as discussed in the introduction.
The model consists of continuity equations for two charged species,
namely electrons and positive ions with particle densities $n_e$ and $n_+$
\begin{eqnarray}
\label{1}
\partial_t\;n_e \;+\; \partial_X J_e
&=& source ~,
\\
\label{2}
\partial_t\;n_+ \;+\; \partial_X J_+
&=& source ~. \end{eqnarray}
Their space charges can modify the externally applied field $E$
through the Poisson equation
\begin{equation}
\label{3}
\partial_X E = {{\rm e}\over{\varepsilon_0}} \;(n_+ -n_e)~. 
\end{equation}
In the simplest approximation, diffusion is neglected and particle 
current densities $J_e$ and $J_+$ are approximated by drift only
\be
\label{4}
 J_e = -n_e \;\mu_e \; E ~~~,~~~
 J_+ =  n_+ \;\mu_+ \; E  ~,
\ee
where the drift velocity here is assumed to be linearly dependent
on the local field with mobilities $\mu_+\ll\mu_e$.

Two ionization processes are taken into account: 
the $\alpha$ process of ionization by electron impact in the
bulk of the gas, and the $\gamma$ process of electron emission
by ion impact onto the cathode. 
In a local field approximation, the $\alpha$ process
is modeled as a local source term in the continuity equations
\be
source = | J_e | \;\bar\alpha(|E|)~~~,~~~
\bar\alpha(|E|)=Ap\;\alpha\left(\frac{|E|}{Bp}\right)~,
\ee
where $p$ is the pressure of the gas. (The mobilities then scale
with inverse pressure $\mu_e=\bar\mu_e/p$ and $\mu_+=\bar\mu_+/p$.)
In the classical Townsend approximation \cite{Raizer}, the function
\be
\label{alphadef}
\alpha({\cal E})=e^{{\textstyle -(1/|{\cal E}|)}^{\textstyle s}}
\ee
is characterized by the single parameter $s$ with typical values
$s=1/2$ or 1 depending on the type of gas. Our numerical results 
are for the most common value $s=1$.

The parameter $\gamma$ is the probability that a positive ion hitting
the cathode leads to emission of a free electron into the gas.
For a discharge of length $d$ with the anode at $X=0$ and the cathode
at $X=d$, the $\gamma$ process enters as boundary condition at $X=d$
\be
\label{G6}
|J_e(d,t)|=\gamma \;|J_+(d,t)|~,
\ee
while ions are absent at the anode
\be
\label{G5}
J_+(0,t)=0~.
\ee
The electric potential $U$ between the electrodes is
\be 
\label{defU}
U(t)=\Phi(0,t)-\Phi(d,t)>0~~~,~~~E(X,t)=-\partial_X\Phi~.
\ee
With this convention, the average electric field $E$ is positive.
Equations (\ref{1})-(\ref{defU}) define the classical model.

\subsection{Reformulation and dimensional analysis}

For the further calculation, it is useful to note, that
the continuity equations (\ref{1}), (\ref{2}) together with the 
Poisson equation (\ref{3}) in one dimension result in the spatial
conservation of the total electric current
\be
\varepsilon_0\partial_t E+{\rm e}\left(J_+-J_e\right)=J(t)
~~~,~~~\partial_XJ=0~,
\ee
and that the ion current density $J_+$ (\ref{4}) with the help of (\ref{3})
can be completely expressed by $J_e$ and $E$
\be
J_+
= \frac{\mu_+}{\mu_e}
\left(-J_e+\frac{\varepsilon_0}{\rm e}\mu_eE\;\partial_XE\right)~.
\ee

By dimensional analysis, the independent dimensionless parameters
of the model are identified. It is convenient to introduce 
the following dimensionless times, lengths and fields
\ba
x=\frac{X}{X_0}~~~&,&~~~\tau=\frac{t}{t_0}~, \nn \\
\sigma(x,\tau)=\frac{n_e(X,t)}{n_0} ~~~&,&~~~
{\cal E}(x,\tau)=\frac{ E(X,t)}{E_0}~, 
\ea
where
\begin{eqnarray}
X_0&=&\frac1{Ap}~~,~~E_0=Bp~,~~
\\
\frac{X_0}{t_0}&=&\mu_eE_0=\bar\mu_eB\;p^0~~,~~
n_0=\frac{\epsilon_0E_0}{{\rm e}X_0}=\frac{\epsilon_0\;AB}{\rm e}\;p^2~.
\nonumber
\end{eqnarray}
The equations now take the form
\ba
\label{g1}
\partial_\tau\sigma&=&\partial_x j_e
+j_e \alpha({\cal E})~,\\
\label{g2}
\partial_\tau {\cal E}&=&j(\tau) -(1+\mu) j_e 
- \mu {\cal E} \partial_x {\cal E}
~,
\ea
where $j_e= \sigma{\cal E}=-{\rm e}J_e/[{\rm e}n_0\;X_0/t_0]$ 
is the dimensionless conductive current carried by the electrons, 
\be
j=\frac{J}{{\rm e}n_0\;X_0/t_0}\propto\frac{J}{p^2}~~~\mbox{and}~~~
u=\frac{U}{E_0X_0}\propto\frac{U}{p^0}
\ee
are the dimensionless total current and potential, and
\be
\mu=\frac{\mu_+}{\mu_e}\propto p^0~~~\mbox{and}~~~L=\frac{d}{X_0}=A\;pd
\ee
are the ratio of ion over electron mobility and the dimensionless 
length of the gas discharge layer.

We here have also recalled the scaling properties with pressure
$p$, such that the pressure dependent similarity laws easily can be
identified in the dimensionless results below.

\subsection{The stationary problem}

For a given dimensionless total current $j$, mobility ratio $\mu$, 
secondary emission coefficient $\gamma$, functional form 
$\alpha({\cal E})$ as in (\ref{alphadef}) and dimensionless system length 
$L$, the stationary solutions of (\ref{g1}), (\ref{g2}) are determined by
\ba
\label{h1}
d_xj_e&=& - \alpha({\cal E})\; j_e~,\\
\label{h2}
\mu {\cal E} d_x {\cal E}&=&j-(1+\mu) j_e~,
\ea
together with the boundary conditions (\ref{G6}), (\ref{G5})
that conveniently are expressed by $j$ as
\ba
\label{h3}
j_e(0)=j ~~~\mbox{and}~~~ j_e(L)=j\;e^{-L_\gamma}
\ea
with
\ba
\label{h3g}
L_\gamma=\ln\frac{1+\gamma}{\gamma}~.
\ea
We assume that $\alpha({\cal E})>0$ and $\partial\alpha/\partial|{\cal E}|>0$
within the relevant range of fields ${\cal E}$. We prove in Appendix A 
that this determines a unique solution for the two functions $j_e(x)$ 
and ${\cal E}(x)$. Finally, the integrated field yields the potential
\be
\label{h4}
u=\int_0^L {\cal E}(x)\;dx~,
\ee
and hence the current-voltage-characteristics $u(j)$. 

\subsection{A global conservation law}

$\alpha({\cal E}(x))$ for all solutions $(j_e(x),{\cal E}(x))$ 
is related to $L_\gamma$ through the global conservation law 
\be
\label{constr}
\int_0^L \alpha({\cal E}(x))\;dx=L_\gamma~.
\ee
This can be seen by formally integrating Eq.\ (\ref{h1})
with the boundary condition $j_e(0)=j$ with the result
\be
j_e(x)=j\;e^{-\int_0^x \alpha({\cal E}(x'))\;dx'}~,
\ee
and by evaluating this solution with the boundary condition 
$j_e(L)=j\;e^{-L_\gamma}$ at $L$.
The identity (\ref{constr}) also can be found in \cite{Engel,Raizer}. 

It follows immediately that for a bounded function with 
$\alpha({\cal E})\le 1$ for all ${\cal E}$ as in (\ref{alphadef}), 
the system size $L$ needs to be larger than $L_\gamma$
\be
L\ge L_\gamma
\ee
to sustain a stationary self-sustained discharge. This is true 
for arbitrary currents $j$ and space charge effects.

The identity (\ref{constr}) also plays a prominent role 
in the small current expansion about the Townsend limit,
as we will see now.

\section{Analytical small current expansion}

\subsection{The Townsend limit}

The well-known Townsend limit can be understood as a consequence 
of (\ref{constr}):
for currents $j$ so small that $\partial_x{\cal E}\approx 0$ 
in (\ref{h2}), the electric field is constant ${\cal E}(x)={\cal E}_T$.
Eq.\ (\ref{constr}) then reduces to the familiar ``ignition condition''
\cite{Raizer}
\be
\label{Town}
\alpha({\cal E}_T)L=L_\gamma~~~\Longleftrightarrow~~~
\gamma\;\left(e^{\;\alpha({\cal E}_T)L}-1\right)=1~.
\ee
The Paschen curve relates the potential $u_T={\cal E}_TL$
in the Townsend limit to the system size $L$
through $\alpha(u_T/L)=L_\gamma/L$. In particular,
for the form of Eq.\ (\ref{alphadef}), the Paschen curve is
\be
\label{Paschen}
u_T(L,\gamma)=\frac{L}{\ln^{1/s}(L/L_\gamma)}~,
\ee
while the field is
\be
\label{ET}
{\cal E}_T(L,\gamma)=\frac1{\ln^{1/s}(L/L_\gamma)}~.
\ee
In dimensionless form, $u_T$ and ${\cal E}_T$  depend only 
on the secondary emission coefficient $\gamma$, 
system size $L$ and the parameter $s$ in (\ref{alphadef}).
The Townsend field ${\cal E}_T$ increases monotonically
with decreasing system size $L$ and diverges for 
$L\downarrow L_\gamma$. The Paschen curve $u_T(L,\gamma)$ 
(\ref{Paschen}) has a minimum at $L=L_\gamma\;e^{1/s}$ 
and diverges both for $L\downarrow L_\gamma$ and for 
$L\to \infty$.

\subsection{The argument of Engel and Steenbeck}

In the old German textbook of Engel and Steenbeck \cite{Engel}, 
the following argument for an expansion about the Townsend limit
can be found: write the electric field as the Townsend field
${\cal E}_T$ plus a perturbation $\Delta(x)$, and note that 
the potential is the integrated field:
\ba
{\cal E}(x)={\cal E}_T+\Delta (x)~~~,~~~
u=u_T+\int_0^L\Delta(x)\;dx~.
\ea
The local impact ionization coefficient can then be expanded about
$\alpha({\cal E}_T)$ as
\be
\alpha({\cal E}(x))=\alpha({\cal E}_T)+\alpha'({\cal E}_T)\;\Delta(x)
+\frac{\alpha''({\cal E}_T)}{2}\;\Delta^2(x)+\ldots~.
\ee
For fixed system size $L$ and parameter $L_\gamma$, 
the global constraint (\ref{constr}) relates different solutions 
${\cal E}(x)$ to $\alpha({\cal E}_T)L$ through
\ba
\alpha({\cal E}_T)L&=&\int_0^L\alpha({\cal E}(x))\;dx
\nonumber\\
&=&\alpha({\cal E}_T)L+\alpha'({\cal E}_T)\;\int_0^L\Delta(x)\;dx
\nonumber\\
&&+\frac{\alpha''({\cal E}_T)}{2}\;\int_0^L\Delta^2(x)\;dx+\ldots~,
\ea
where the expansion of $\alpha$ was used in the second step.
This identity allows to express $\int_0^L\Delta(x)\;dx$
by the higher order terms $\int_0^L\Delta^n(x)\;dx$, $n=2,3,\ldots$.
Insertion of this expansion into the definition of $u$ yields
\be
\label{Eng}
u=u_T- \frac{\alpha''({\cal E}_T)}{2\;\alpha'({\cal E}_T)}\;
\int_0^L\Delta^2(x)\;dx
+\ldots
\ee
Since $\Delta^2$ is positive and since $\alpha$ is assumed to be 
an increasing function of ${\cal E}$, the sign of the correction
is determined by the sign of $\alpha''$. This statement from
\cite{Engel} is recalled in the recent literature
\cite{Raizer,Kolobov}. It should be noted that the estimate (\ref{Eng}) 
is valid as long as $\left| \alpha^{(n)}\int\Delta^ndx \right| 
\ll \left|\alpha''\int\Delta^2 dx\right|$ for all $n\ge3$.

The question is now how to calculate $\int_0^L \Delta^2(x) dx$.
In \cite{Engel}, a result is quoted referring to a long calculation
without reference. The same result is given more than 60 years later 
in \cite{Raizer} in Section 8.3 with a sketch of an argument 
and again without reference. The argument assumes that 
$|J_+|\gg|J_e|$ throughout the discharge volume. This assumption
is in disagreement with the boundary condition (\ref{G5}).
A somewhat different argument based on a constant space charge
through the whole system is given in \cite{Klyar57}.
In Ref.\ \cite{Raizer}, the electric field profile is assumed
to be ${\cal E}(x)\propto \sqrt{1-x/x_0}$, while \cite{Klyar57}
it is assumed to be ${\cal E}(x)\propto\left(1-x/x_0\right)$ 
where the length scale $x_0$ depends on the current $j$. In both cases,
the breakdown of the approximation is determined from the field 
vanishing at the anode: ${\cal E}(L)\approx 0$. This prescription 
yields no dependence on $\gamma$ at all, quite in contrast to our 
results below. The functional forms for ${\cal E}(x)$ should be 
compared with our systematic analytical results (\ref{ansatzE}), 
(\ref{solE1}) below (note that we reversed the order of anode and 
cathode), and with our numerically derived field profiles 
in Figs.\ 4 and 5. They do not justify the ans\"atze given above.

Rather a consistent ansatz is chosen in \cite{Petro97},
and the structure of their expansion in terms of $e^{-L_\gamma}$ 
and $L_\gamma$ is quite similar to ours below.
However, these authors fail to incorporate the global
conservation law (\ref{constr}), and get a correction
already in linear order, in contrast to the rigorous 
result (\ref{Eng}) above.

\subsection{A systematic expansion in small $j$}

We now perform a systematic expansion in powers of $j$ about the
Townsend limit. In principle, this expansion can be extended to 
arbitrary order. We have evaluated it up to $O(j^3)$. We write
the field correction as a power series in $j$, namely $\Delta(x)=
j\;{\cal E}_1(x)+j^2\;{\cal E}_2(x)+\ldots$, and use the same 
ansatz for the current $j_e(x)$
\ba
\label{ansatzE}
{\cal E}(x)=&{\cal E}_T+&j\;{\cal E}_1(x)+j^2\;{\cal E}_2(x)+\ldots~,\\
\label{ansatzj}
j_e(x)=&&j\;\iota_1(x)+j^2\;\iota_2(x)+\ldots~,
\ea
and we introduce the short hand notation
\be
\alpha=\alpha({\cal E}_T)~~,~~
\alpha'=\alpha'({\cal E}_T)~~,~~
\alpha''=\alpha''({\cal E}_T)~,~\ldots
\ee
for the Taylor expansion of 
\ba
\alpha({\cal E}(x))&=&\alpha
+\alpha'\left(j{\cal E}_1(x)+j^2{\cal E}_2(x)+\ldots\right)
\nn \\
&&~~+\;\frac{\alpha''}{2}\left(j{\cal E}_1(x)+j^2{\cal E}_2(x)
+\ldots\right)^2
+\ldots
\ea
Insertion of the ans\"atze (\ref{ansatzE}), (\ref{ansatzj})
into Eqs.\ (\ref{h1}) and ordering in powers of $j$ yields
\ba
\label{i1}
O(j^1)~:~~ && \partial_x\iota_1(x)=-\iota_1(x)\;\alpha~, \\
\label{i2}
O(j^2)~:~~&& \partial_x\iota_2(x)=-\iota_2(x)\;\alpha
-\iota_1(x)\;\alpha'{\cal E}_1(x)~,~\ldots
\ea
For Eq.\ (\ref{h2}), the same procedure gives
\ba
\label{E0}
O(j^0)~:~~&& \partial_x{\cal E}_T=0~,\\
\label{E1}
O(j^1)~:~~&& \mu{\cal E}_T\partial_x{\cal E}_1=1-(1+\mu)\iota_1(x)~,\\
\label{E2}
O(j^2)~:~~&& \mu{\cal E}_T\partial_x{\cal E}_2
+\mu{\cal E}_1\partial_x{\cal E}_1=-(1+\mu)\iota_2(x),~\ldots
\ea
The boundary condition (\ref{h3}) at the anode ($x=0$) yields
\be
\label{bc0}
\iota_1(0)=1~~~,~~~\iota_2(0)=0~~~,~~~\iota_3(0)=0~~~,~~~\ldots
\ee
The boundary condition (\ref{h3}) at the cathode ($x=L$) most 
conveniently is evaluated with the help of the global conservation law 
(\ref{constr}). Taking into account that $L_\gamma$ is independent 
of $j$, the expanded form reads
\ba
\label{alpha0}
O(j^0)~:~~ && \alpha\;L=L_\gamma~,\\
\label{alpha1}
O(j^1)~:~~ && \int_0^L {\cal E}_1(x)\;dx=0~,\\
\label{alpha2}
O(j^2)~:~~ && \int_0^L \left(\alpha'\;{\cal E}_2(x)
              +\alpha''\;\frac{{\cal E}_1^2(x)}{2}\right)\;dx=0~,\\
\label{alpha3}
O(j^3)~:~~ && \int_0^L \left(\alpha'\;{\cal E}_3
              +\alpha''\;{\cal E}_1{\cal E}_2
              +\alpha'''\frac{{\cal E}_1^3}{3!}\right)\;dx=0~,\nn \\
&& \ldots
\ea
where the first equation (\ref{alpha0}) reproduces the ignition condition
(\ref{Town}). Finally, the potential $u$ from (\ref{h4}) is
\ba
\label{Uexp}
u&=&u_T(L,\gamma)+j\int_0^L {\cal E}_1(x)\;dx
  +j^2\int_0^L {\cal E}_2(x)\;dx \nn \\
 &&~  +\;j^3\int_0^L {\cal E}_3(x)\;dx+\ldots
\ea
The lowest order $u_T(L,\gamma)$ reproduces the Paschen curve (\ref{Paschen}).
Eq.\ (\ref{alpha1}) reveals immediately that the order
$j^1$ in $u$ has to be absent.
For the order $j^2$ in (\ref{Uexp}), the function ${\cal E}_1(x)$
has to be calculated. First, 
\be
\label{iota1}
\iota_1(x)=e^{-\alpha x}
\ee
is the solution of (\ref{i1}) and (\ref{bc0}). $\iota_1(x)$ has to be 
inserted into (\ref{E1}) which now can be solved analytically
up to a constant of integration. This constant is determined 
by (\ref{alpha1}). The result is
\be
\label{solE1}
{\cal E}_1(x)=
\frac{\alpha x-\frac{L_\gamma}{2}+
(1+\mu)\left(e^{-\alpha x}-\frac{1-e^{-L_\gamma}}{L_\gamma}\right)
      }{\alpha\mu{\cal E}_T}~.
\ee
For the contribution in order $j^2$ to the potential, the calculation
of ${\cal E}_1$ is sufficient since with the help of (\ref{alpha2}):
\ba
\int_0^L {\cal E}_2(x)\;dx&=&
-\;\frac{\alpha''}{2\;\alpha'}\int_0^L {\cal E}_1^2(x)\;dx 
= -\;\frac{\alpha''}{2\;\alpha'}\;
\frac{F(\gamma,\mu)}{\alpha^3\mu^2{\cal E}_T^2}~, \nn
\ea
with the function
\ba
\label{F}
F(\gamma,\mu)&=&\frac{L_\gamma^3}{12}
+(1+\mu)\left(2-L_\gamma-2e^{-L_\gamma}-L_\gamma 
e^{-L_\gamma}\right)
\nonumber\\
&&+\;(1+\mu)^2\left(\frac{1-e^{-2L_\gamma}}{2}
               -\frac{(1-e^{-L_\gamma})^2}{L_\gamma}\right)~.
\ea
The function is plotted in Fig.\ 1. 
Within the interesting parameter regime, it depends strongly
on $\gamma$ and invisibly on $\mu$. Here we use the parameter
range for $\gamma$ suggested by \cite{Raizer} and the maximal 
mobility ratio $\mu=\mu_+/\mu_e=0.0095$ is reached for 
the lightest molecules, namely hydrogen.

\begin{figure}[h]
\setlength{\unitlength}{0.9cm}
\begin{picture}(6,6)
\epsfxsize=7cm
\epsfbox{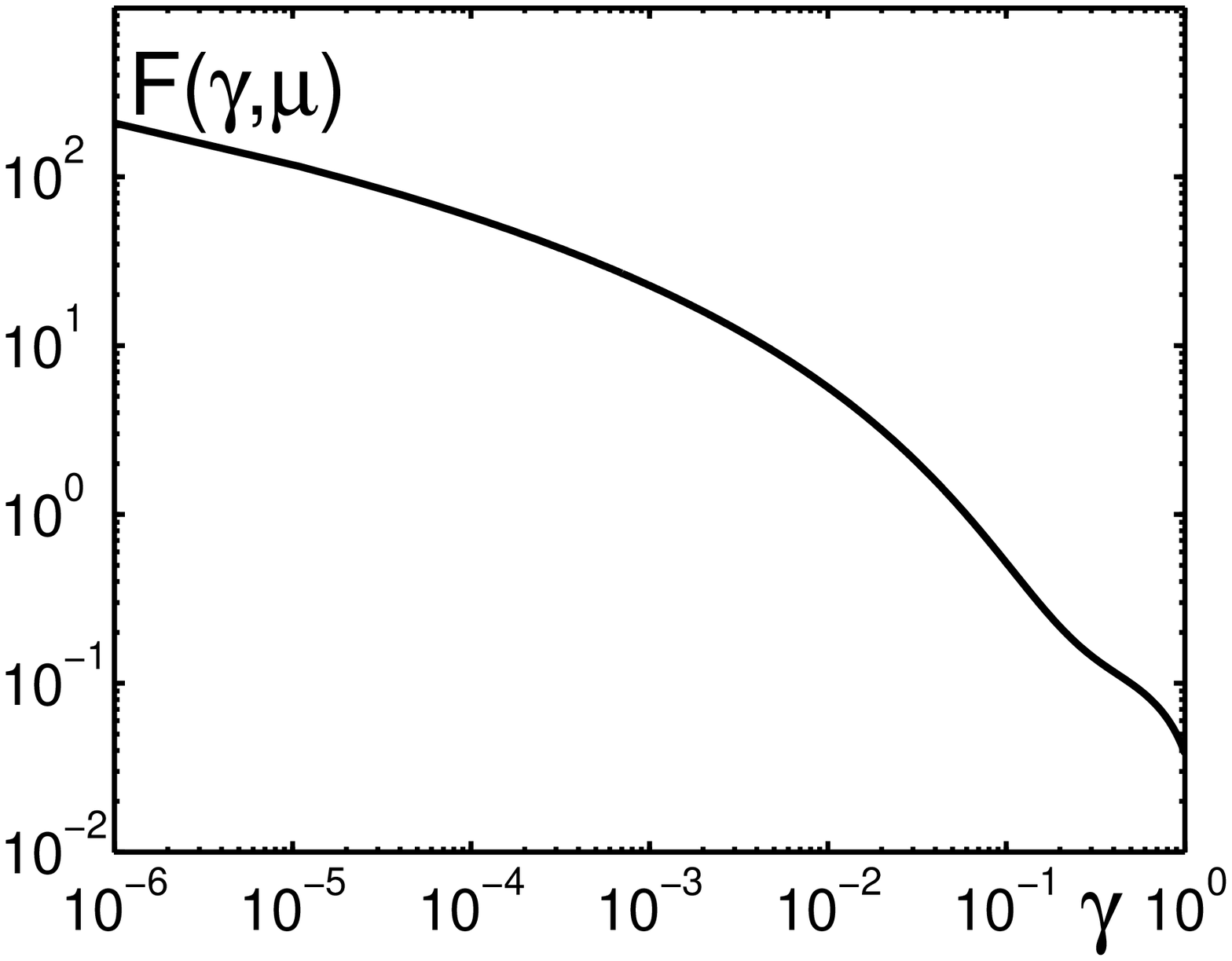}
\end{picture}
\begin{center}
\begin{minipage}{9cm}
\small FIG.\ 1.
Plot of $F(\gamma,\mu)$ as a function of $\gamma$ 
in a double-logarithmic plot. The dependence on $\mu$
for realistic values $0\le\mu\le0.0095$ is too weak
to be visible in the plot. However, $F(\gamma,\mu)$
varies over almost 4 orders of magnitude as a function of $\gamma$.
\end{minipage}
\end{center}
\end{figure}

The small current expansion of the current-voltage-characteristics
is in this approximation
\be
\label{U}
u=u_T-\left(\frac{j}{\mu}\right)^2\;\frac{{\cal E}_T\;\alpha''}{2\;\alpha'}\;
\frac{F(\gamma,\mu)}{\left(\alpha{\cal E}_T\right)^3}+O(j^3)~.
\ee
The range of validity of this expansion can be easily estimated
by inserting (\ref{solE1}) into (\ref{ansatzE}): the correction
to the field due to the current should not exceed half of the Townsend
field, so
\be
\label{range}
j\stackrel{<}{\sim}\frac{{\cal E}_T}
{\displaystyle 2\;\max_x\left|{\cal E}_1(x)\right|}
\stackrel{\gamma\le1}{=}\frac{{\cal E}_T}{2\;{\cal E}_1(L)}
\stackrel{L_\gamma\gg1}{\approx}\frac{\mu{\cal E}_T^2}{L}~.
\ee

In view of the very good fit of this expansion with our numerical
results to be presented below in Fig.\ 6, and in view of the 
interesting bending structure of the numerically derived current-voltage
characteristics in Fig.\ 9 below, it seemed promising to calculate 
the next term of the expansion of order $j^3$
\be
\label{U3}
u=u_T-\left(\frac{j}{\mu}\right)^2\;\frac{{\cal E}_T\;\alpha''}{2\;\alpha'}\;
\frac{F(\gamma,\mu)}{\left(\alpha{\cal E}_T\right)^3}
+\left(\frac{j}{\mu}\right)^3 f_3+O(j^4)~.
\ee
The function $f_3$ can also be calculated fully analytically 
and along the same lines: first $\iota_2(x)$ 
is derived from (\ref{i2}) and (\ref{bc0}) and inserted into 
the o.d.e.\ (\ref{E2}) for ${\cal E}_2(x)$. The equation is solved,
and the constant of integration is determined by (\ref{alpha2}).
Then $\int_0^L{\cal E}_3(x)\;dx$ is derived from (\ref{alpha3})
and the explicit expressions for ${\cal E}_1(x)$ and ${\cal E}_2(x)$
and inserted into (\ref{Uexp}). However, the result of this
calculation is still considerably longer than (\ref{F})
and does not show any simple structure. We therefore 
will not give the explicit form here. Rather, we summarize essential
steps of the calculation in more detail in Appendix B.
The final steps are done by computer algebra (mathematica).
Some of the results for $f_3$ are shown in Fig.\ 6 and compared
with numerical solutions of the full problem (\ref{h1}) -- (\ref{h4}).

\subsection{Discussion of the result}

Translating back from dimensionless to physical units,
the result reads
\ba
U(J)&=&U_T-\left(\frac{J}{\varepsilon_0\mu_+}\right)^2\;
\left.\frac{E \partial_E^2\bar\alpha}{2\;\partial_E \bar\alpha}\right|_{E_T}\;
\frac{F(\gamma,\mu)}{(\bar\alpha \;E_T)^3}
\nonumber\\
&&~~+\left(\frac{J}{\epsilon_0\mu_+}\right)^3 \frac{f_3}{(E_0^2 Ap)^3}
+\ldots~, \nn
\ea
with the dimensionless coefficient
\be
\left.\frac{E \partial_E^2\bar\alpha}{2\;\partial_E \bar\alpha}\right|_{E_T}
=\frac{{\cal E}_T\;\alpha''}{2\;\alpha'}
=\frac{sE_0^s- (s+1) E_T^s}{2 E_T^s}~.
\ee
The coefficient of $J^2$ changes sign for 
${\cal E}_T=E_T/E_0=[s/(s+1)]^{1/s}$. With the help of (\ref{ET}), 
this transition at $\alpha''=0$ can be located on the Paschen curve; 
it occurs at 
\be
\label{Lcrit}
L_{crit}=L_\gamma\;e^{\:1+1/s}~.
\ee
This is always on the right branch of the Paschen curve,
since the minimum is at $L=L_\gamma\;e^{1/s}$.

The result agrees qualitatively with the one given by Raizer \cite{Raizer} 
and Engel and Steenbeck \cite{Engel}. In particular, the leading
order correction is also of order $\alpha''(j/\mu)^2$.
However, the explicit coefficient of $j^2$ differs:
while the coefficient in \cite{Engel,Raizer} does not
depend on $\gamma$ at all, we find that the dependence on $\gamma$ 
is essential, as the plot of $F$ in Fig.\ 1 clearly indicates.
In fact, within the relevant range of $10^{-6}\le \gamma\le 10^0$, 
this coefficient varies by almost four orders of magnitude. 
We remark that it indeed would be quite a surprising mathematical result
if the Townsend limit itself would depend on $\gamma$ as in (\ref{ET}), 
but the small current expansion about it would not.
Our $\gamma$-dependent analytical result 
also excellently fits our numerical solutions, 
as we will show in the next section.

\section{Numerical solutions}

We now discuss our numerical results for the voltage $u$ as a function 
of total current $j$, secondary emission coefficient $\gamma$, mobility ratio 
$\mu$ and system size $L$, as resulting from Eqs.\ (\ref{h1})--(\ref{h4}).
We will work with the Townsend approximation 
$\alpha({\cal E})=e^{-1/|{\cal E}|}$ (\ref{alphadef}) with $s=1$ 
as the standard case \cite{Raizer}.

\subsection{The numerical method}

In Appendix A, we showed that the solution $u=u(j)$ is unique for
fixed $\gamma$, $\mu$ and $L$, and we proved the useful property 
that the system size $L$ is a monotonically decreasing function
of the electric field ${\cal E}(0)$ at the anode,
$d L/d{\cal E}(0)<0$ (\ref{A3}), for fixed $\gamma$, $\mu$ and $j$. 
The second observation lays the basis for our numerical iteration procedure:

First the two o.d.e.'s (\ref{h1}), (\ref{h2}) are integrated from $x=0$ 
with the known initial value $j_e(0)=j$ and some guessed initial value 
${\cal E}(0)$ towards larger $x$. The equations are
integrated until for some $x=\bar x$, we find the value
$j_e(\bar x)=j\;e^{-L_\gamma}$ that should be assumed at the fixed
system size $x=L$. If $\bar x>L$, a larger value of ${\cal E}(0)$
is chosen for the next iteration step, and if $\bar x<L$, a smaller
${\cal E}(0)$ where a linear interpolation of $d\bar x/d{\cal E}(0)$ is used.
This iteration loop is continued until the boundary condition (\ref{h3}) 
at $L$ is obeyed with sufficient accuracy. The potential $\phi(x)$
is integrated together with $j_e(x)$ and ${\cal E}(x)$ by adding
the third o.d.e.\ $\partial_x\phi=-{\cal E}$\cite{Manuel}. The voltage 
$u$ over the system is $u=\phi(0)-\phi(L)$. 

For the numerical integration of the o.d.e.'s, we used the lsodar.f 
routine of the ODEPACK package from the free-ware site netlib.org.
It integrates initial value problems for sets of first order o.d.e.'s 
and chooses automatically the appropriate numerical method for stiff 
or non-stiff systems. At the same time, 
it locates the roots of any specified function. We defined this
function as $j_e(x)-j\;e^{-L_\gamma}$ which returns the value
$\bar x$ for the next iteration loop with high precision.

\subsection{Parameters $L$, $\gamma$ and $\mu$, and $j/\mu$-scaling}

The problem depends on the following parameters:
the first one is the system size $L$ which is proportional 
to $pd$ in physical units. It can take arbitrary values; 
we explore a continuous range of $L$ on both the left and the right 
branch of the Paschen curve.

The second parameter is the secondary emission coefficient $\gamma$ which is
determined by both the gas and the cathode surface. Increasing $\gamma$
decreases the minimum breakdown voltage which is $e L_\gamma$ as discussed
after Eq.\ (\ref{Lcrit}). This mechanism can be used 
for improving performance in technical applications like plasma
display panels \cite{pdp}. According to \cite{Raizer},
$\gamma$ can take values between $10^{-6}$ and $10^{-1}$, in extreme
cases even larger. We show results either for the two extreme cases
$10^{-6}$ and $10^{-1}$, or we show one representative result for 
$\gamma=10^{-2}$.

The third parameter is the mobility ratio $\mu=\mu_+/\mu_e$ 
of the charged species. Since ions are much heavier than electrons,
$\mu$ is always much smaller than 1. The largest value of $\mu=0.0095$
\cite{Raizer} is reached for the lightest molecules, namely hydrogen.
As a standard, we use the value $\mu=0.0035$ for nitrogen.

The functional form of the defining equations (\ref{h1})-(\ref{h3g}) 
and of the small current expansion (\ref{U3}) suggest that $u$ 
in leading order does not depend on $j$ and $\mu$ separately,
but only on the scaling variable $j/\mu$ and on the factor $(1+\mu)\approx1$. 
This observation motivates our choice of the variable $j/\mu$ in the 
following figures. However, Fig.\ 7 will show that for large $j/\mu$ 
in the abnormal glow regime and for large systems $L$, there is some small
$\mu$-dependent correction to this scaling behavior. Reconsidering
(\ref{h1})-(\ref{h3g}), this means that the factor $(1+\mu)$ can not
simply be equated with 1 even for $\mu<10^{-2}$, but yields some 
correction. In physical terms, the substitution of $(1+\mu)$ by 1
means the elimination of the field increase in the anode fall region,
and we conclude that in large systems in the glow regime, the anode
fall yields some small contribution to the current-voltage-characteristics.

\subsection{General features of the current-voltage-characteristics}

We now give an overview over our numerical results
in the full parameter regime of the current-voltage-characteristics 
$u(j)$ from Townsend up to abnormal glow discharge as a function
of rescaled current $j/\mu$, system size $L$ and secondary emission 
coefficient $\gamma$.
In Figs.\ 2 and 3, we plot $u$ as a function of $j/\mu$ and $L$
for $\gamma=10^{-6}$ and $\gamma=10^{-1}$, respectively. 
The plots follow the style of an experimental plot in \cite{Klyar66},
which is reproduced as Fig.\ 1 in \cite{Kolobov}. To the best of our
knowledge, our Figs.\ 2 and 3 for the first time present numerical results 
in the same style.

Comparing the two figures for different $\gamma$, it can be noted
that on the one hand, the shapes look qualitatively similar,
while on the other hand, the actual parameter regimes of potentials,
currents and system sizes vary by an order of magnitude or more.
Let us now consider the common features.

In the limit of small current $j$ (i.e., in the foreground of the figures), 
the curves saturate to a plateau
value which actually reproduces the Paschen curve $u=u_T(L,\gamma)$
from Eq.\ (\ref{Paschen}).

\begin{figure}[h]
\setlength{\unitlength}{1cm}
\begin{picture}(8,6)
\epsfxsize=8.5cm
\epsfbox{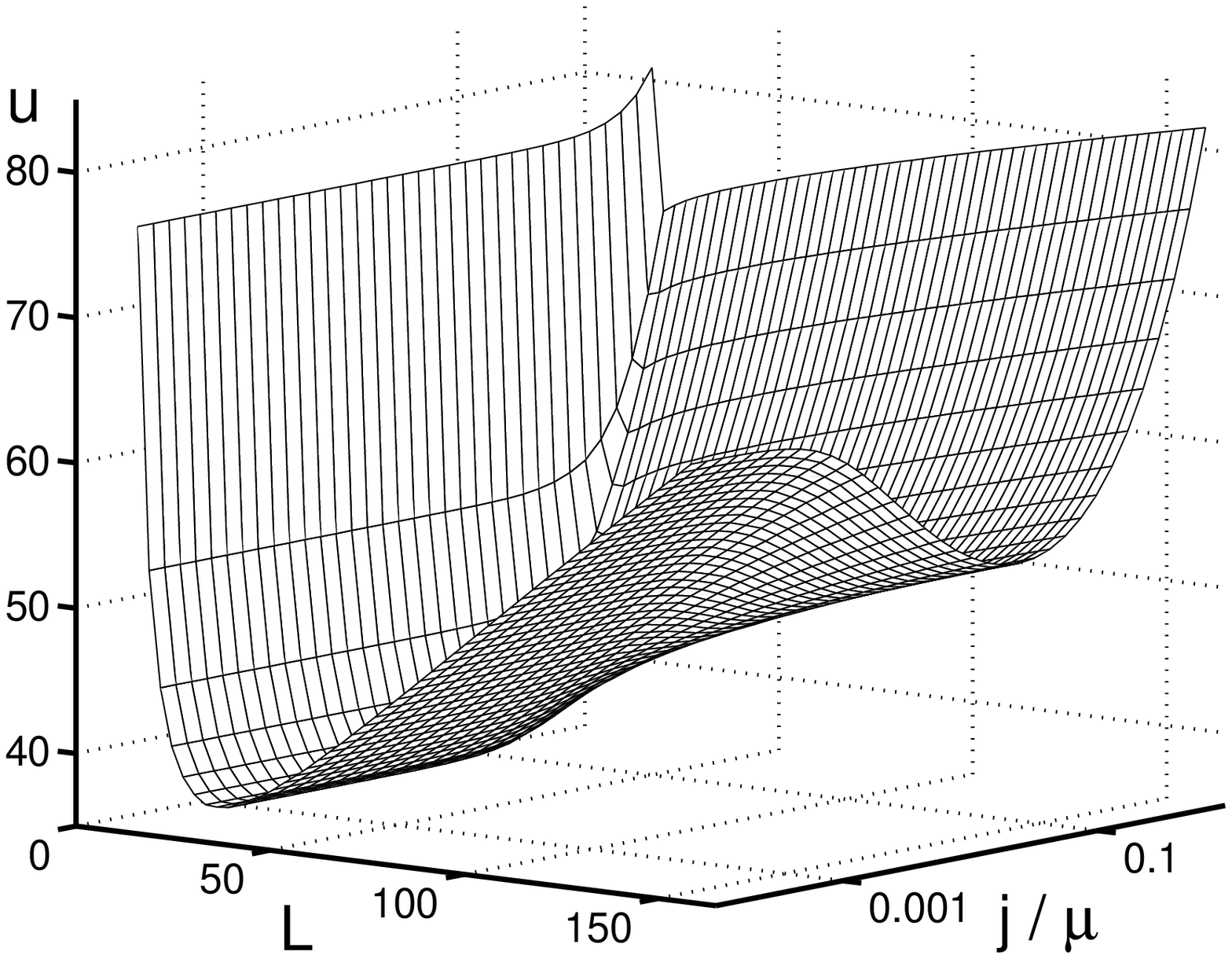}
\end{picture}
\begin{center}
\begin{minipage}{8cm}
\small FIG.\ 2.
$u$ as a function of $j/\mu$ and $L$ for the small secondary emission 
coefficient $\gamma = 10^{-6}$. The parameter range is 
$3 \cdot 10^{-7}/\mu \leq j/\mu \leq 5 \cdot 10^{-3}/\mu$ for
$\mu=0.0035$ and $17.3 \leq L \leq 160$.
\end{minipage}
\end{center}
\end{figure}

\begin{figure}[h]
\setlength{\unitlength}{1cm}
\begin{picture}(8,5.5)
\epsfxsize=7.8cm
\epsfbox{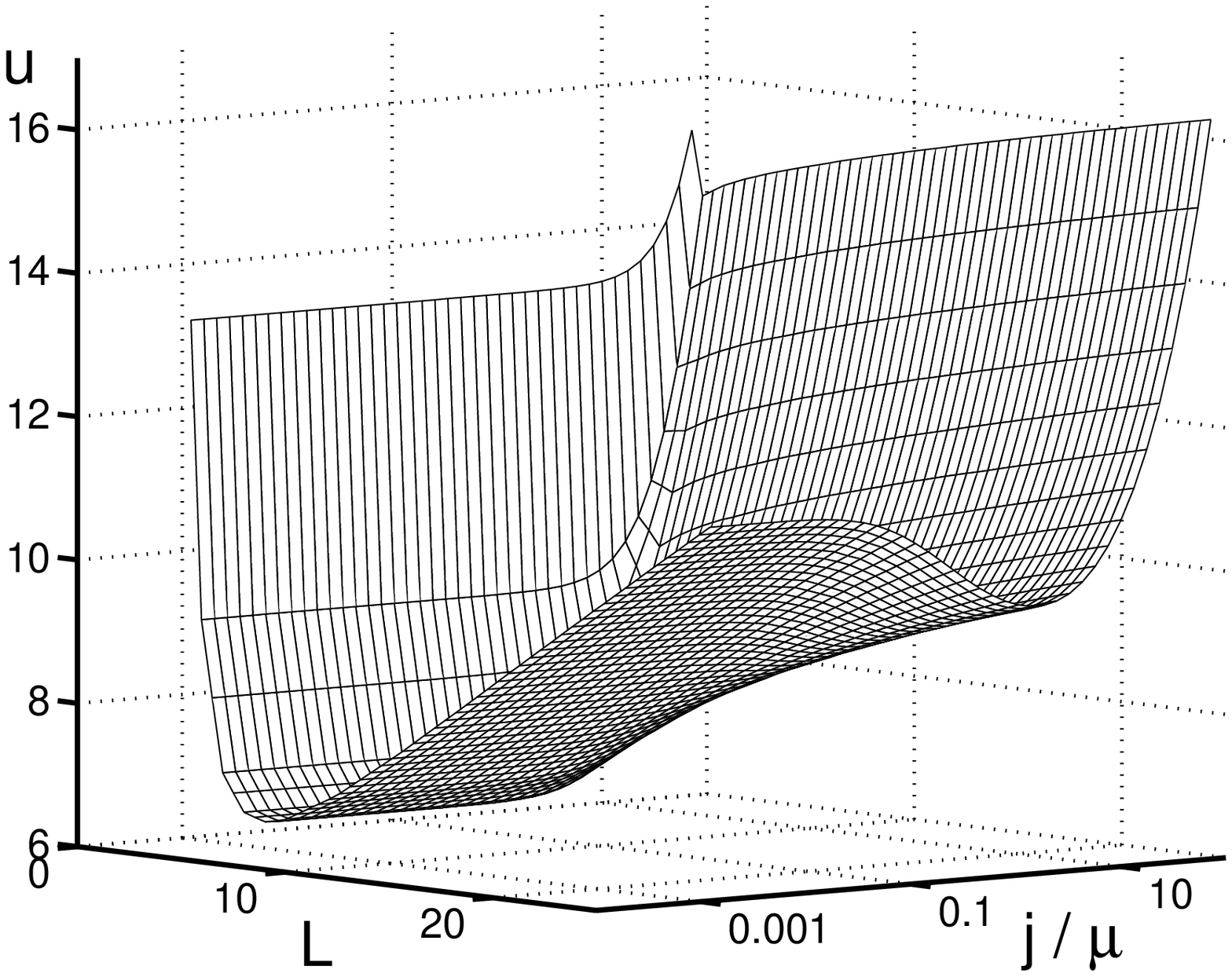}
\end{picture}
\begin{center}
\begin{minipage}{8cm}
\small FIG.\ 3.
Plot as in Fig.\ 2, but now for $\gamma=10^{-1}$.
The parameter range is $ 10^{-6}/\mu \leq j/\mu \leq 7 \cdot 10^{-2}/\mu$ 
for $\mu=0.0035$ and $3 \leq L \leq 28$.
\end{minipage}
\end{center}
\end{figure}

Following $u=u(j)$ along a line of fixed system size $L$,
we get the current-voltage-characteristics for this particular
system characterized by the two parameters $L$ and $\gamma$. 
For these curves $u=u(j)$, two features can be noted. 

First, for larger $L$, the voltage $u$ first decreases for increasing 
current $j$. This is the familiar Townsend-to-glow-transition 
with negative differential conductivity. For larger $j$, the voltage 
$u$ increases again towards the regime of abnormal glow. However, 
in the minimum of the potential, there is no plateau in contrast to 
experimental plots. This is because we solve the purely one-dimensional 
system 
without the possibility of a lateral growth of the glow discharge column. 

Second, for smaller values of $L$, in particular, when starting
from the left branch of the Paschen curve, the voltage does not decrease
for increasing current, but it increases immediately. We will discuss
this different bifurcation structure in more detail at the end of this
section.

\subsection{Spatial profiles}

It is instructive to study the spatial profiles of electron
current $j_e(x)$ and field ${\cal E}(x)$ for different system sizes.
In Figs.\ 4 and 5, we plot such profiles for $L=e L_\gamma$ 
and for $L=e^3 L_\gamma=e L_{crit}$. The smaller
system size $e L_\gamma$ coincides with the minimum of the Paschen
curve, while $L_{crit}=e^2 L_\gamma$ (\ref{Lcrit}) is the system size 
where $\alpha''$ in (\ref{U3}) changes sign. So for $L<L_{crit}$
the voltage increases initially in the small current expansion
around the Townsend limit, while for  $L>L_{crit}$ it decreases.

Note that in contrast to previous plots, e.g., in \cite{Raizer},
our cathode is on the right hand site at $x=L$, because we found
it more convenient to work with a positive field ${\cal E}$. 
The electron current is normalized by the total current.

In each plot, the profiles for the two smallest current values
are well described by the small current expansion from Section III.
This is in agreement with the range of validity of these expansions
of $j/\mu \alt 0.08$ for $L=eL_\gamma=L_{crit}/e$ or 
of $j/\mu \alt 1.3\cdot 10^{-3}$ for $L=eL_{crit}$, resp., 
estimated according to (\ref{range}).

\begin{figure}[h]
\setlength{\unitlength}{1cm}
\begin{picture}(8,7)
\epsfxsize=8.5cm
\epsfbox{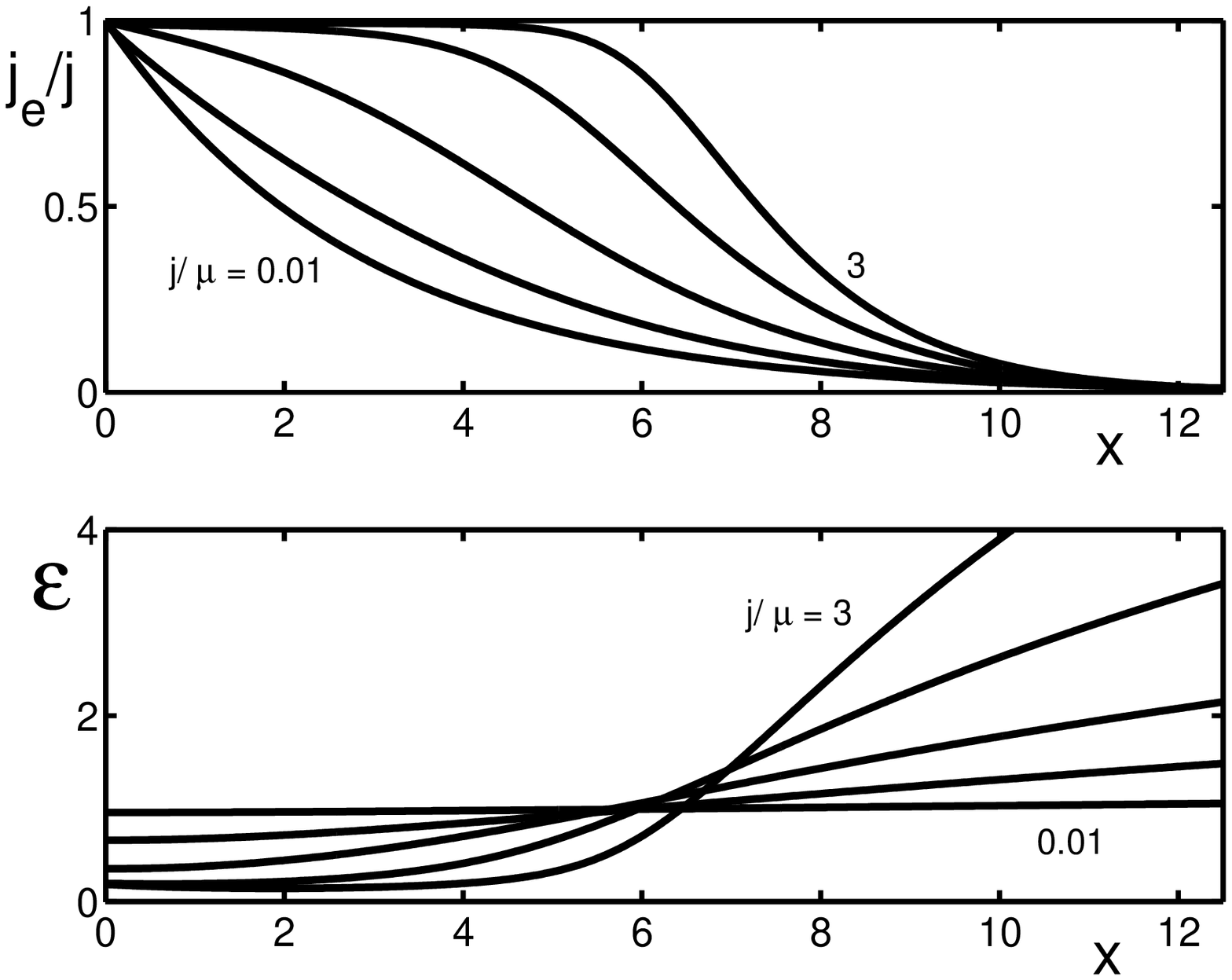}
\end{picture}
\begin{center}
\begin{minipage}{8cm}
\small FIG.\ 4.
Spatial profiles $j_e(x)/j$ and ${\cal E}(x)$ for system size 
$L=e L_\gamma$ at the minimum of the Paschen curve.
Plotted are curves for $j/ \mu = 0.01$, 0.1, 0.3, 1, 3. 
Other parameters: $\gamma=0.01$, $\mu=0.0035$.
\end{minipage}
\end{center}
\end{figure}

\begin{figure}[h]
\setlength{\unitlength}{1cm}
\begin{picture}(8,7)
\epsfxsize=8.5cm
\epsfbox{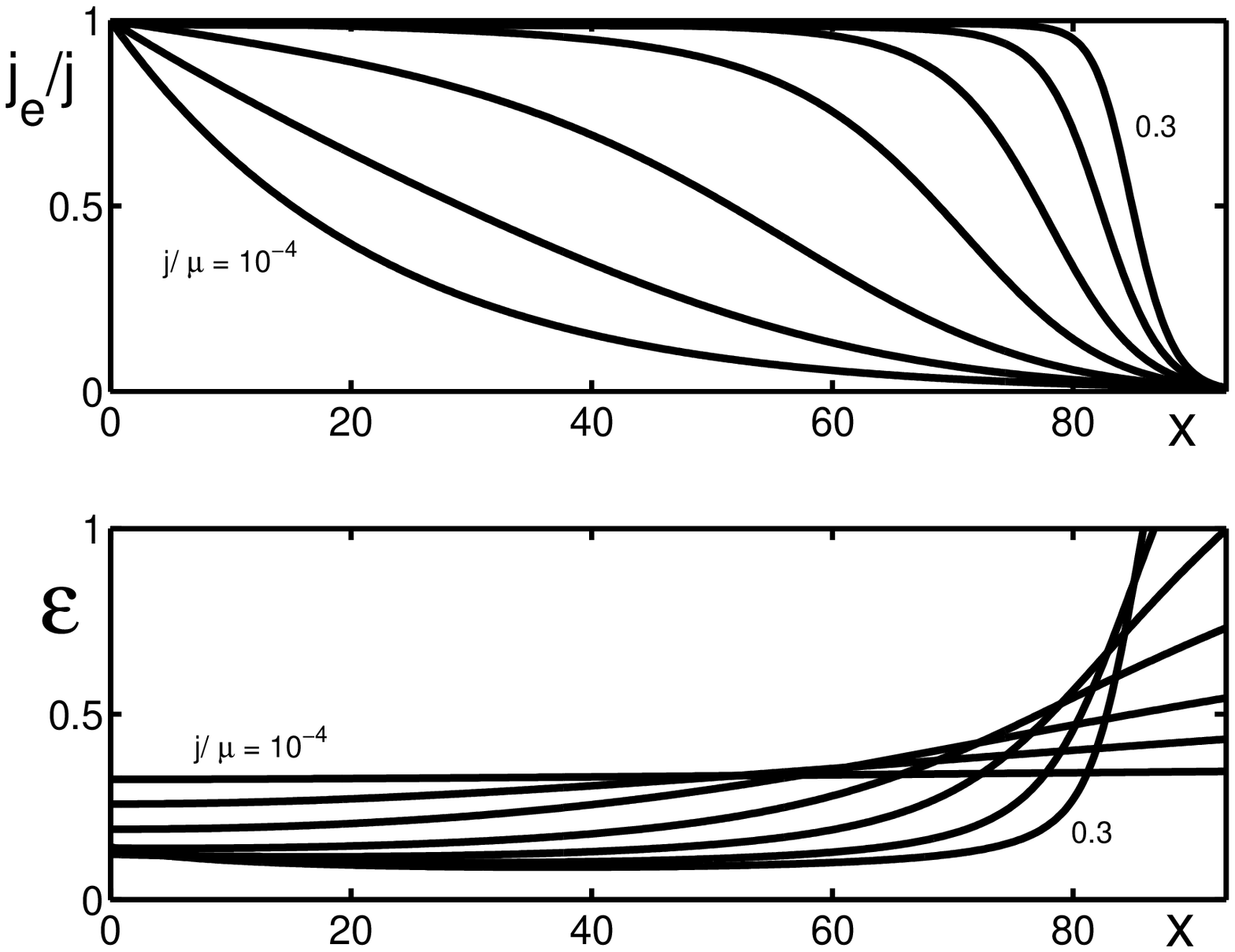}
\end{picture}
\begin{center}
\begin{minipage}{8cm}
\small FIG.\ 5.
The same as in the previous figure, but now for the larger system size
$L=e^3L_\gamma$. The current now explores the smaller values 
$j/ \mu = 10^{-4}$, $10^{-3}$, $3 \cdot 10^{-3}$, 0.01, 0.03, 0.1, 0.3.
\end{minipage}
\end{center}
\end{figure}

For larger currents, a separation into resistive column on the left
and cathode fall on the right becomes pronounced. While in the smaller
system, both regions take about equal parts, in the larger system,
the cathode fall takes only a small part of the volume on the 
right hand side.

\subsection{Comparison of numerical and analytical results}

Let us now compare the current-voltage-characteristics
corresponding with these profiles with our analytical results
from Section III. In Fig.\ 6, the numerical results for $u(j)$ are
plotted as a thick solid line, and the analytical expansions (\ref{U}) 
and (\ref{U3}) up to second or third order in $j/\mu$ as thin solid 
and dashed lines, respectively. For the calculation of the third order, 
the procedure described in Appendix A has been followed.
Fig.\ 6 shows that in particular the expansion up to order $(j/\mu)^3$
gives a very good agreement at least within the range of validity
of $j/\mu \alt 0.08$ or $1.3\cdot 10^{-3}$, resp., according to 
(\ref{range}). 

\begin{figure}[h]
\setlength{\unitlength}{1cm}
\begin{picture}(8,7)
\epsfxsize=8.5cm
\epsfbox{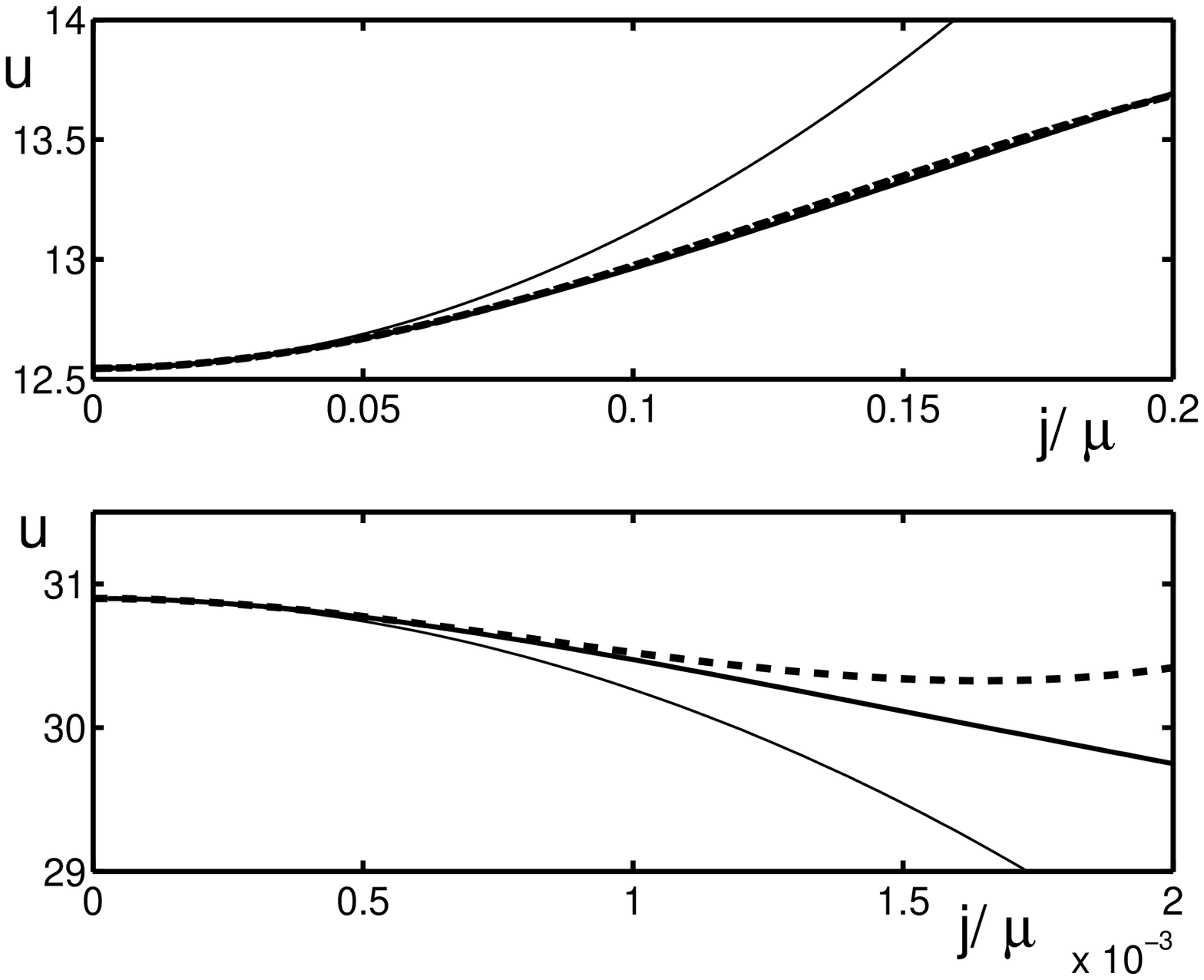}
\end{picture}
\begin{center}
\begin{minipage}{8cm}
\small FIG.\ 6.
$u(j)$ for the systems from Figs.\ 4 and 5 in the small current limit:
upper plot $L=eL_\gamma$, lower plot $L=e^3L_\gamma$, 
both with $\gamma=0.01$.
Numerical result (thick solid line), analytical result (\ref{U}) up
to second order in $j/\mu$ (thin solid line) and analytical result
(\ref{U3}) up to third order in $j/\mu$ (dashed line).
\end{minipage}
\end{center}
\end{figure}

\subsection{Corrections to $j/\mu$-scaling}

Fig.\ 7 shows the current-voltage-characteristics for the same two systems,
but now up to larger values of the current than in Fig.\ 6. 
Actually, the same current range is explored in each system 
as in the corresponding Figs.\ 4 and 5.

In addition, in Fig.\ 7 we test the $j/\mu$-scaling by plotting $u$ 
as a function of $j/\mu$ within the physical range of $\mu$-values, 
including the limit of $\mu=0$. It can be noted that for short systems 
or small currents, the $\mu$-correction is negligible; this means that 
$(1+\mu)$ can be replaced by 1 in Eq.\ (\ref{h2}) without visible
consequences. In contrast, for large systems and large currents,
there is a small, but visible $\mu$-correction to the dominant
$j/\mu$-scaling.

\begin{figure}[h]
\setlength{\unitlength}{1cm}
\begin{picture}(8,7)
\epsfxsize=8.5cm
\epsfbox{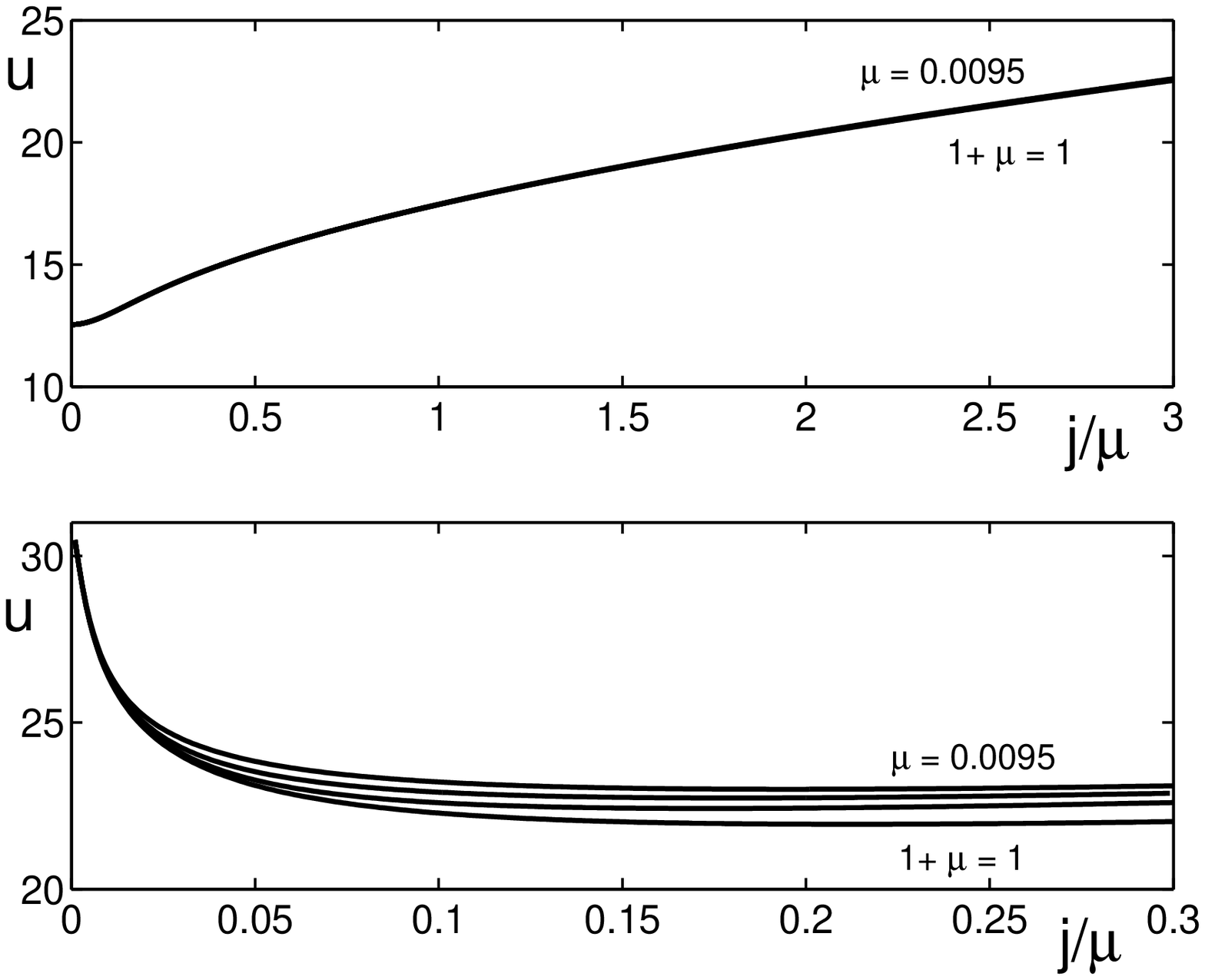}
\end{picture}
\begin{center}
\begin{minipage}{8cm}
\small FIG.\ 7.
Current-voltage characteristics for the same two systems as in
Figs.\ 4, 5, and 6 (upper plot $L=eL_\gamma$, lower plot $L=e^3L_\gamma$), 
but now for a larger current range than in Fig.\ 6.
Furthermore, besides the curves for the nitrogen value for the
mobility ratio $\mu = 0.0035$, also the curves for $\mu= 0.0095$
(hydrogen), $\mu = 0.001$ and the limiting value $\mu=0$ are shown.
\end{minipage}
\end{center}
\end{figure}

\subsection{Discussion of bifurcation structures}

We now set the final step in the quantitative understanding
of the current-voltage-characteristics at the transition
from Townsend to glow discharge. We characterize the transitions
as subcritical, mixed or supercritical and locate them in
parameter space.

Fig.\ 8 gives an overview over the different behaviors for 
$\gamma=0.01$. It corresponds to different $L$-sections of plots 
as in Figs.\ 2 and 3, but now with $j/\mu$ plotted on a linear 
rather than a logarithmic scale.
For the terminology of sub- or supercritical bifurcations,
it should be noted that $j/\mu=0$ is a solution for arbitrary $u$. 
So the complete $u$-axis is a solution, too. 

In the case of $L=0.85 L_{crit}$,
there is a pure forward or supercritical bifurcation:
$u$ increases monotonically as $j/\mu$ increases.
In contrast, for $L=1.05 L_{crit}$, the bifurcation
is purely subcritical: with increasing $j/\mu$, the voltage
first decreases, and eventually it increases again.
This subcritical behavior continues down to $L=L_{crit}=L_\gamma e^2$ 
where $\alpha''$ changes sign. So indeed, the sign change of 
$\alpha''$ in the small current expansion determines the transition
from subcritical to some other behavior. However, for $L<L_{crit}$,
the system not immediately enters the supercritical regime,
but some unexpected ``mixed'' behavior can be seen:
for increasing $j/\mu$, the voltage $u$ first increases,
then it decreases and then it increases again.
We distinguish ``mix$_{\rm I}$'' where the voltage minimum 
at finite $j/\mu$ is smaller than the Townsend voltage,
and ``mix$_{\rm II}$'' where it is larger.

\begin{figure}[h]
\setlength{\unitlength}{1cm}
\begin{picture}(8,6.2)
\epsfxsize=8cm
\epsfbox{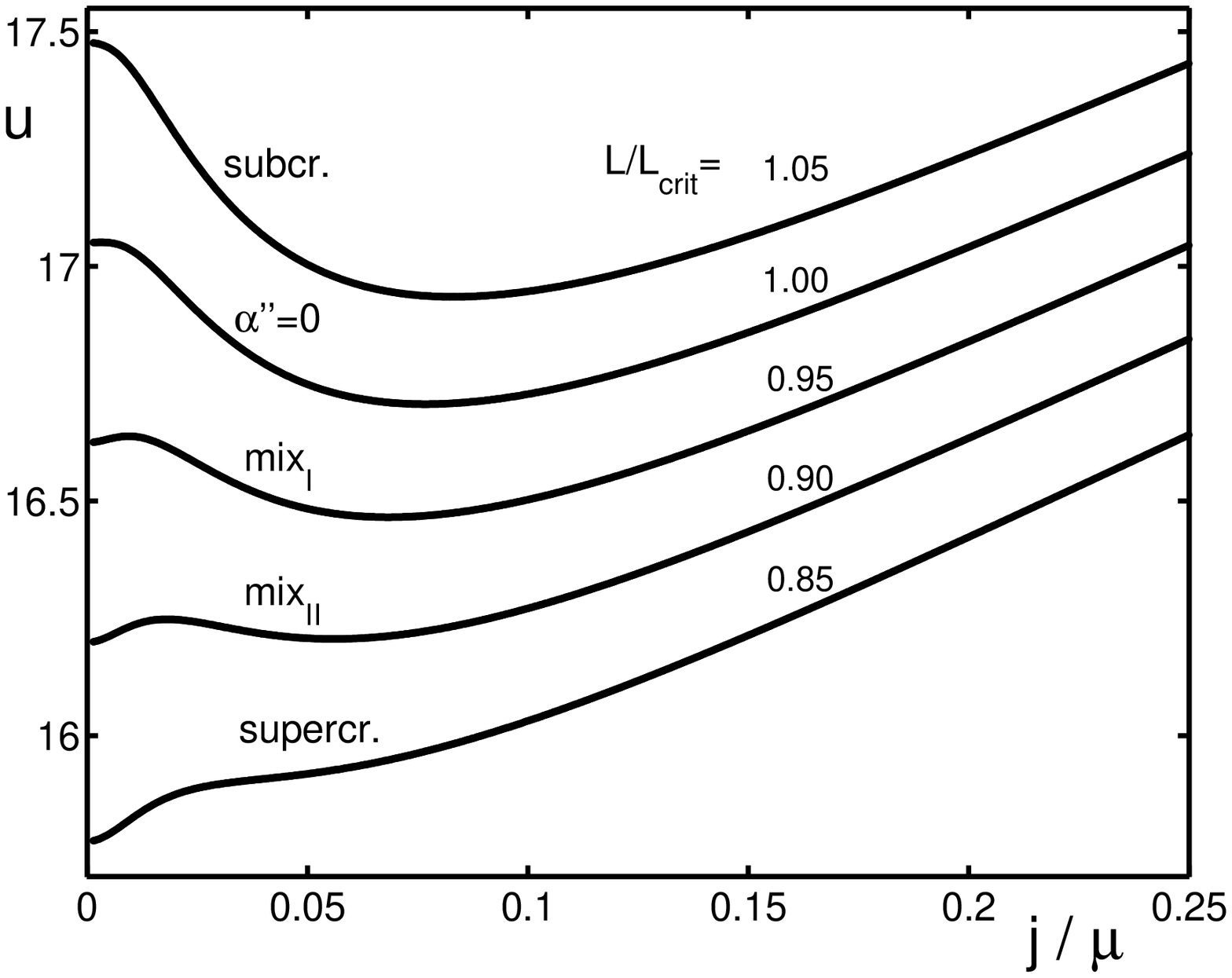}
\end{picture}
\begin{center}
\begin{minipage}{8cm}
\small FIG.\ 8.
The current-voltage-characteristics for fixed parameters
$\gamma=0.01$ and $\mu=0.0035$ and different system sizes,
measured in multiples of $L_{crit}=L_\gamma e^2$.
Shown are all possible bifurcation structures from supercritical
up to the familiar subcritical case for various values of $L$.
\end{minipage}
\end{center}
\end{figure}

Fig.\ 9 shows a zoom into Fig.\ 8: a smaller range of current
and of system sizes. The form of an upwards parabola of the order
$(j/\mu)^2$ next to the $u$-axis is well described by the analytical 
small current expansion of Section III. However, in contrast to 
initial hopes, the turn-over of the curves at $j/\mu\approx 0.02$
is not covered by the small current expansion up to order
$(j/\mu)^3$ whose coefficient even changes sign within the parameter
range of Fig.\ 9. In fact, the range of validity of the expansion
breaks down at $j/\mu\le 0.017$, just briefly before the first
interesting bending structure in the characteristics.

\begin{figure}[h]
\setlength{\unitlength}{1cm}
\begin{picture}(8,6)
\epsfxsize=8cm
\epsfbox{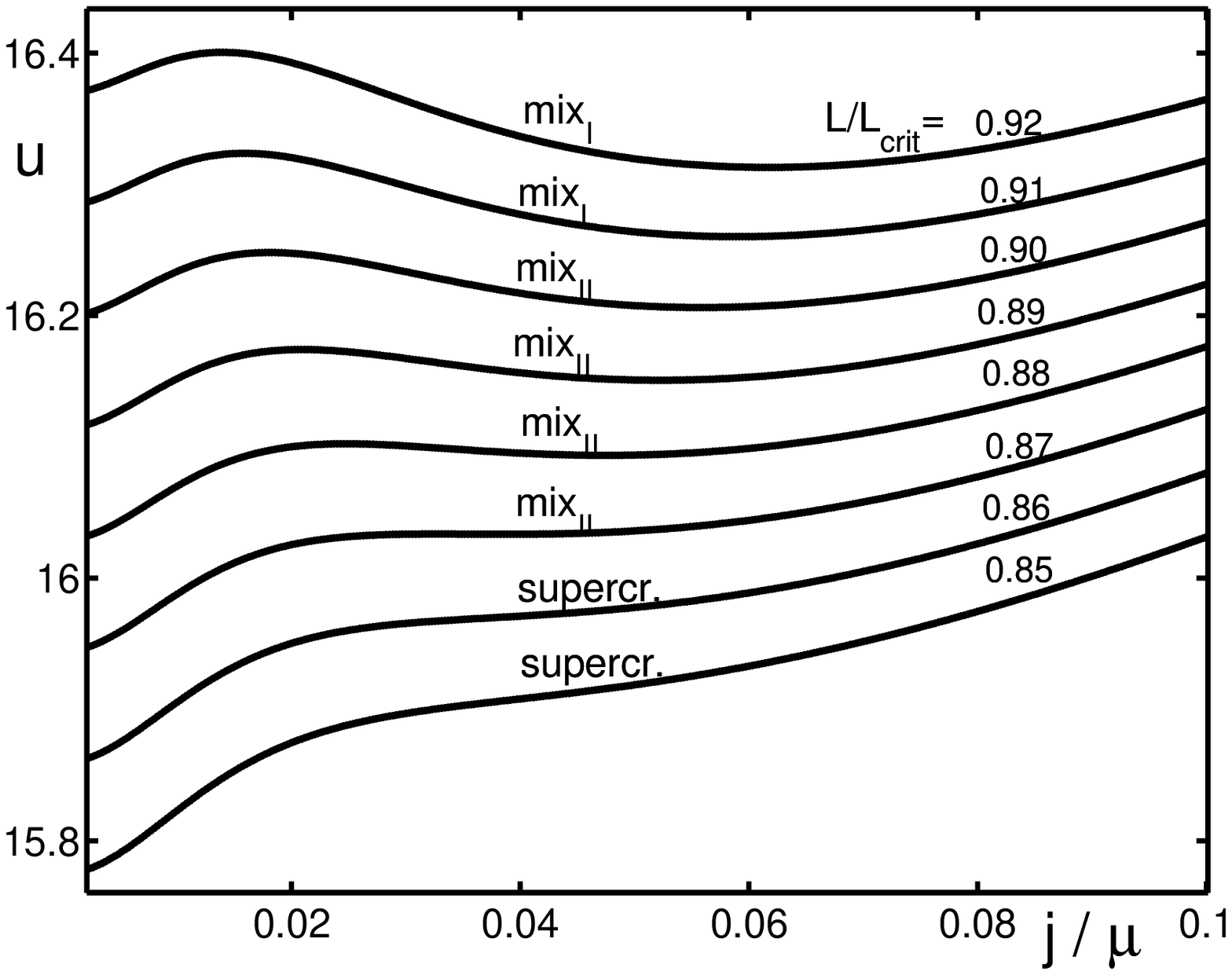}
\end{picture}
\begin{center}
\begin{minipage}{8cm}
\small FIG.\ 9.
Zoom into Fig.\ 8 with smaller values of $j/\mu$ and system sizes $L$.
The range of validity of the analytical small current expansion (\ref{U3})
is $j/\mu\le 0.017$.
\end{minipage}
\end{center}
\end{figure}

Finally, in Fig.\ 10, the bifurcation behavior in the full 
parameter range of $\gamma$ is explored. The transition from subcritical
to mix$_{\rm I}$ always takes place when $\alpha''$ changes sign,
i.e., at system size $L_{crit}=L_\gamma e^2$.
The transitions from mix$_{\rm I}$ to mix$_{\rm II}$ and then
further to supercritical occur at smaller relative system sizes 
$L/L_{crit}$ when the secondary emission coefficient $\gamma$ is smaller.
All transitions occur on the right branch of the Paschen
curve, since its minimum is at $L/L_{crit}=e^{-1}=0.368$.

\begin{figure}[h]
\setlength{\unitlength}{1cm}
\begin{picture}(8,6.5)
\epsfxsize=8.5cm
\epsfbox{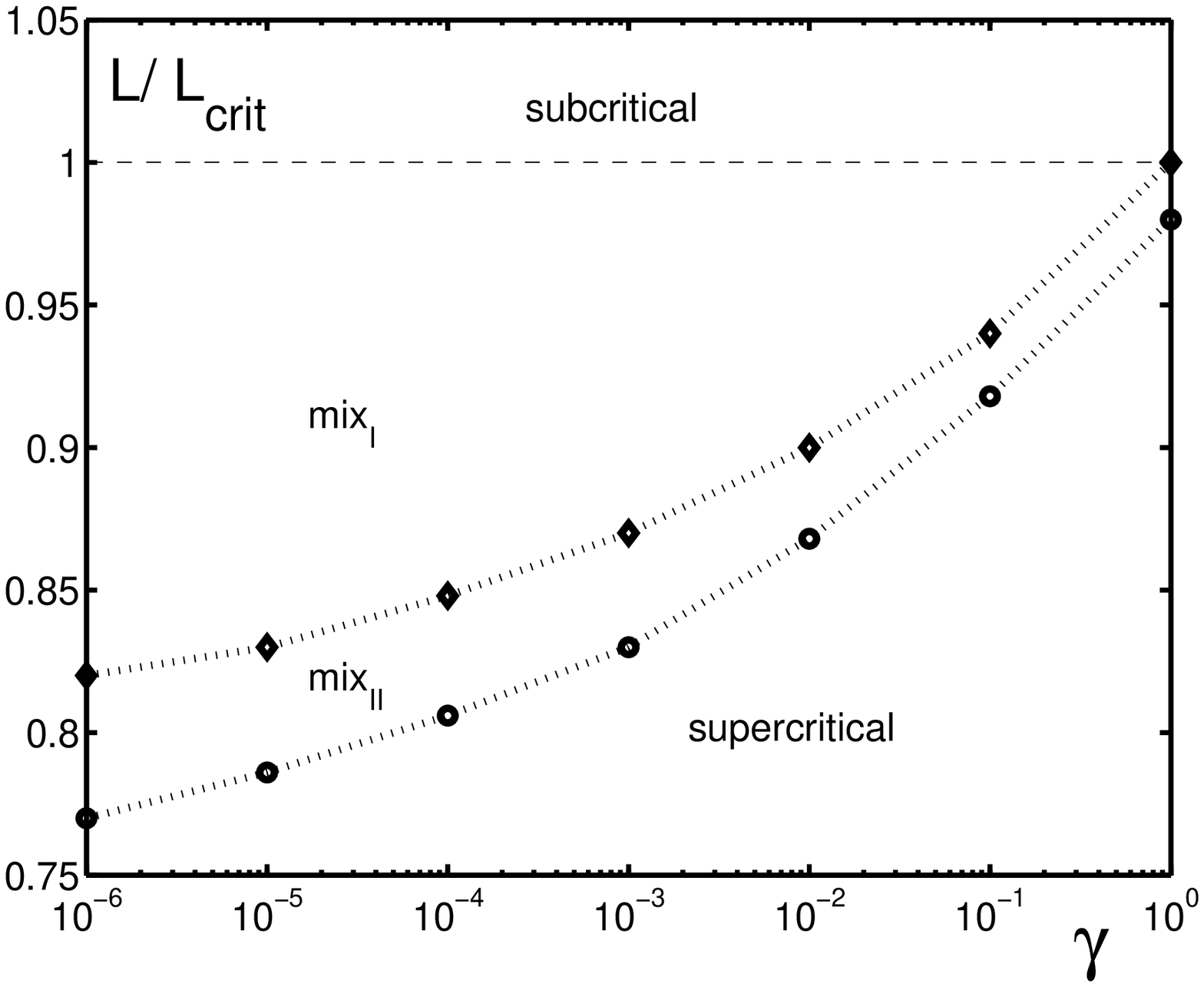}
\end{picture}
\begin{center}
\begin{minipage}{8cm}
\small FIG.\ 10.
Complete overview over the bifurcation behavior of the current-voltage
characteristics as a function of secondary emission coefficient 
$\gamma$ and system size $L$.
\end{minipage}
\end{center}
\end{figure}

\section{Summary and Outlook}

We have studied the classical minimal model that creates a Townsend
or glow discharge, in one-dimensional approximation. The dimensionless
current-voltage characteristics $u=u(j/\mu)$ depends on essentially
only two parameters, the secondary emission coefficient $\gamma$
and the dimensionless system size $L\propto pd$.
(With $j/\mu$-scaling, the further dependence on the small mobility
ratio $\mu=\mu_+/\mu_e$ is very weak and becomes visible only for 
long systems in the normal and abnormal glow regime.)
Numerically, we have fully explored the bifurcation structure as a 
function of $\gamma$ and $L$. Besides the familiar subcritical bifurcation
structure of long systems, for decreasing system size $L$, there
is a sequence of current-voltage characteristics, that we have
called mix$_{\rm I}$ and mix$_{\rm II}$, before the supercritical
transition is reached. This general sequence is the same for all 
relevant values of $\gamma$ while the precise lengths where the 
transitions occur, depend on $\gamma$, cf.\ Fig.\ 10.
Analytically, we have calculated the small current expansion
about the Townsend limit in a systematic expansion.
We found that the term of order $j/\mu$ is missing, and that the
term of order $(j/\mu)^2$ indeed is proportional to the second
derivative of the Townsend coefficient $\alpha''$ according to
the old argument from Engel and Steenbeck \cite{Engel}, but
with a different, strongly $\gamma$-dependent proportionality
constant. We also have calculated the term of order $(j/\mu)^3$.
These analytical expansions are in very good agreement with our
numerical results within their predicted range of validity.

Of course, the study of this minimal model can only be a first step, 
and it has been suggested to include a number of additional features.
First, $\gamma$ might not be constant; while the dependence 
$\gamma=\gamma(I,V)$ \cite{Phelps93III} on global parameters
seems unphysical, the local dependence $\gamma=\gamma(E/p)$
has been suggested \cite{PhelpsPRE97} and experimentally tested 
\cite{Auday}. Second, the particle mobilities might be field-dependent
$\mu_\pm=\mu_\pm(E)$. Third, diffusion was neglected, the approach
was fully local, and only one ion type was considered.
However, our aim was first to settle the predictions of the 
classical model in full parameter space as a corner stone 
and starting point for any future extension that simultaneously
also will increase the number of parameters.

The motivation for this work is the impressive variety of 
spatio-temporal patterns formed in short barrier discharges
\cite{Muenster1,Muenster2,StripeM,Muenster3,HexM,Astrov,Zigstr,Str,filStr}.
The nonlinear element responsible for the spontaneous pattern formation
is believed to be a gas discharge in the parameter range
of the present work. Parameter regions with negative differential
conductivity (NDC) are generally believed to play a decisive role 
in the formation of the instabilities. Knowledge about NDC regions 
and the bifurcation structure in the range of these experiments 
therefore are a condition for their future investigation, and
conversely, properties of the current-voltage characteristics
might be deduced from temporal oscillations or current constrictions.
These pattern formation processes will be subject of our future studies.

\vspace{0.5cm}

{\bf Acknowledgment}: D.S.\ is supported by the 
Dutch organization for Fundamental Research of Matter, FOM.

\begin{appendix}

\section{Uniqueness of the solution of the boundary value problem}

Here we prove that the boundary value problem defined
by (\ref{h1}) -- (\ref{h3g}) for fixed $j$, $\mu$, $\gamma$ and $L$
defines a unique solution $\big(j_e(x),{\cal E}(x)\big)$ and
hence a unique potential $u=\int_0^L {\cal E}(x)dx$. This lays
the ground for our analytical as well as for our numerical procedure.
We will keep $j$, $\mu$ and $\gamma$ fixed within this
appendix, and will discuss how ${\cal E}(x)$ is determined by $L$
and vice versa.

First Eq.\ (\ref{h1}) for $j_e$ is integrated with initial condition
$j_e(0)=j$ (\ref{h3}) and inserted into (\ref{h2}):
\be
\label{A1}
\mu{\cal E}\partial_x{\cal E}
=j\left(1-(1+\mu)\;e^{-\int_0^x \alpha({\cal E}(x))\;dx}\right)
\ee
The boundary condition (\ref{h3}) at $L$ amounts to 
\be
\label{A2}
\int_0^L \alpha({\cal E}(x))\;dx=L_\gamma~,
\ee
where we assume that
\be
\label{alp}
\alpha({\cal E})>0~~\mbox{ and }~~\partial\alpha/\partial{\cal E}>0
~~~\mbox{for all }{\cal E}>0~.
\ee
An initial condition ${\cal E}(0)$ defines a unique solution
${\cal E}(x)$ of (\ref{A1}) and hence a unique system size
$L$ through (\ref{A2}). 
We will show below that $L$ is a monotonically decreasing function 
of ${\cal E}(0)$
\be
\label{A3}
d L/d{\cal E}(0)<0~~~\mbox{for fixed }j,\mu,\gamma~.
\ee
This statement has two immediate consequences:
$(i)$ it shows that ${\cal E}(0)$ and therefore also ${\cal E}(x)$
and $u$ are uniquely determined by $L$; and $(ii)$ it lays
the ground for our numerical iteration procedure where ${\cal E}(0)$
is fixed, and the resulting $L$ is calculated and compared to the
true $L$.

Why is statement (\ref{A3}) true?
Compare two solutions ${\cal E}_{1,2}(x)$ and suppose that
${\cal E}_1(x')>{\cal E}_2(x')$ on some interval $0\le x'\le x$.
Then for the difference, we get from (\ref{A1}) that
\begin{eqnarray}
\label{A4}
\lefteqn{\frac{\mu}{2j\;(1+\mu)}
\left(\partial_x{\cal E}_1^2-\partial_x{\cal E}_2^2\right)}
\\
&=&e^{-\int_0^x \alpha({\cal E}_2(x))\;dx}
-e^{-\int_0^x \alpha({\cal E}_1(x))\;dx}\ge0~,
\nonumber
\end{eqnarray}
where the bound $\ldots\ge0$ is a direct consequence of (\ref{alp}).
So when ${\cal E}_1$ is above ${\cal E}_2$ on some interval $0\le x'\le x$,
then at the end of the interval, we have 
$\partial_x{\cal E}_1^2>\partial_x{\cal E}_2^2$, and ${\cal E}_1$ stays
above ${\cal E}_2$. As a consequence
\be 
{\cal E}_1(x)>{\cal E}_2(x)~~\mbox{for all } x\ge 0~~,\mbox{ if }
{\cal E}_1(0)>{\cal E}_2(0)~.
\ee
Inserting this into (\ref{A2}) and using (\ref{alp}), 
statement (\ref{A3}) results.

\section{The correction of $O(j^3)$ about the Townsend limit}

We here sketch the essential elements of the calculation
of the third term $j^3$ of the expansion about the Townsend limit:
first Eq.\ (\ref{i2}) is integrated. With
(\ref{iota1}) for $\iota_1$, (\ref{solE1}) for
${\cal E}_1$, and the boundary condition (\ref{bc0}),
we get 
\ba
\iota_2(x)&
=\displaystyle-\;\frac{\alpha'\;e^{-\alpha x}}{\alpha^2\mu{\cal E}_T}
&\bigg[\frac{(\alpha x)^2}{2}+(1+\mu)(1-e^{-\alpha x})
\nonumber\\
&&-\left.\alpha x\left(\frac{L_\gamma}{2}
+(1+\mu)\frac{1-e^{-\alpha x}}{L_\gamma}\right)\right]
\ea
This result allows us now to integrate ${\cal E}_2(x)$ 
in (\ref{E2}) with the rather lengthy result
\ba
{\cal E}_2(x)&=&\frac1{2\alpha^2\mu^2{\cal E}_T^3}\;\bigg[
(1+\mu)\frac{\alpha'{\cal E}_T}{\alpha}\;\cdot
\nonumber\\
&&~~\bigg\{-(\alpha x)^2 e^{-\alpha x}+(1+\mu)(e^{-2\alpha x}-2e^{-\alpha x})
\nonumber\\
&&~~+2\left(\frac{L_\gamma}{2}-1+(1+\mu)\frac{1-e^{-L_\gamma}}{L_\gamma}\right)
(\alpha x+1) e^{-\alpha x}\bigg\}
\nonumber\\
&&-\alpha x
\left(L_\gamma-\alpha x+2(1+\mu)\frac{1-e^{-L_\gamma}}{L_\gamma}\right)
\nonumber\\
&&+ 2(1+\mu)\left(\frac{L_\gamma}{2}+(1+\mu)\frac{1-L_\gamma}{L_\gamma}
-\alpha x\right)  e^{-\alpha x}
\nonumber\\
&&-(1+\mu)^2e^{-2\alpha x}\bigg] + C
\ea
The constant of integration $C$ is determined through (\ref{alpha2})
\be
\int_0^L{\cal E}_2\;dx=-\;\frac{\alpha''}{2\;\alpha'}\;
\frac{F(\gamma,\mu)}{\alpha^3\mu^2{\cal E}_T^2}.
\ee
Finally, $\int_0^L {\cal E}_3\;dx$ is determined by ${\cal E}_1$ 
and ${\cal E}_2$ through (\ref{alpha3}) as
\be
\int_0^L{\cal E}_3\;dx=-\;\frac{\alpha''}{\alpha'}
\int_0^L{\cal E}_1{\cal E}_2\;dx - \frac{\alpha'''}{3!\;\alpha'}
\int_0^L{\cal E}_1^3\;dx.
\ee
Insertion of the result in Eq.\ (\ref{Uexp}) yields the third order
expansion (\ref{U3}) with an explicit expression for the function $f_3$.

All integrals can be performed analytically. However, the results
are lengthy and exhibit no simplifying structure. 
Therefore, we rather have performed the remaining integrals
by computer algebra. Results are shown in Fig.~6.

\end{appendix}

\end{multicols}


\begin{references}


\bibitem{Jonas1} Piet Jonas, Ph.D. thesis,
{\tt http://www.physik.} {\tt uni-greifswald.de/$\sim$jonas/Thesis/index.html}
(in German), Greifswald 1998.

\bibitem{Jonas2} B. Bruhn, B.-P. Koch, and P. Jonas, Phys. Rev. E
{\bf 58}, 3793-3805 (1998).

\bibitem{Bruhn} B. Bruhn and B.-P. Koch, Phys. Rev. E {\bf 61},
3078-3092 (2000).

\bibitem{Muenster1} 
Yu.A. Astrov, E. Ammelt and H.-G. Purwins,
Phys. Rev. Lett. {\bf 78}, 3129 (1997).

\bibitem{Muenster2} 
Yu.A. Astrov and Y.A. Logvin,
Phys. Rev. Lett. {\bf 79}, 2983-2986 (1997).

\bibitem{StripeM}
E. Ammelt, Yu.A. Astrov and H.-G. Purwins,
Phys. Rev. E {\bf 55}, 6731 (1997).

\bibitem{Muenster3} 
Y.A. Astrov, I. M\"uller, E. Ammelt and H.-G. Purwins,
Phys. Rev. Lett. {\bf 80}, 5341 (1998).

\bibitem{HexM}
E. Ammelt, Yu.A. Astrov and H.-G. Purwins,
Phys. Rev. E {\bf 58}, 7109 (1998).

\bibitem{Astrov} 
L.M. Portsel, Y.A. Astrov, I. Reimann, H,-G. Purwins,
J. Appl. Phys. {\bf 81}, 1077 -- 1068 (1997).

\bibitem{Zigstr}
C. Str\"umpel, Y.A. Astrov, E.Ammelt and  H.-G. Purwins,
Phys. Rev. E {\bf 61}, 4899 (2000).

\bibitem{Str}
C. Str\"umpel, Y.A. Astrov, and H.-G. Purwins,
Phys. Rev. E {\bf 62}, 4889-4897 (2000). 

\bibitem{filStr}
C. Str\"umpel,  H.-G. Purwins, and Y.A. Astrov,
Phys. Rev. E {\bf 63}, 026409 (2001).

\bibitem{Engel} 
A. von Engel and M. Steenbeck, 
{\bf Elektrische Gasentladungen. Ihre Physik und Technik}, 
Vol. II (Springer, Berlin 1934).

\bibitem{Radehaus89} 
H.-G. Purwins, C. Radehaus, T. Dirksmeyer, R. Dohmen,
R. Schmeling, and H. Willebrand,
Phys. Lett. A {\bf 136}, 480 (1989).

\bibitem{Radehaus90}
C. Radehaus, R. Dohmen, H. Willebrand, and F.-J. Niedernostheide,
Phys. Rev. A {\bf 42}, 7426 (1990).

\bibitem{Phelps93I}
Z.L. Petrovic and A.V. Phelps,
Phys. Rev. E {\bf 47}, 2806-2814 (1993).

\bibitem{Phelps93II}
B.M. Jelenkovic, K. R\'ozsa, and A.V. Phelps,
Phys. Rev. E {\bf 47}, 2816-2824 (1993).

\bibitem{Phelps93III}
A.V. Phelps, Z.L. Petrovic, and B.M. Jelenkovic,
Phys. Rev. E {\bf 47}, 2825-2838 (1993).

\bibitem{Petro97} 
Z.L. Petrovic, I. Stefanovic, S. Vrhovac, and J. Zivkovic,
J. Phys. IV France {\bf 7}, Colloque C4, 341-352 (1997).

\bibitem{Schoell}
E. Schoell, {\it Nonequilibrium Phase Transitions in Semiconductors} 
(Springer-Verlag, Berlin Heidelberg, 1987).

\bibitem{Raizer} 
Y.P. Raizer, 
{\bf Gas Discharge Physics}
(Springer, Berlin, 2nd corrected printing, 1997).

\bibitem{Druyve} 
M.J. Druyvesteyn and F.M. Penning,
Rev. Mod. Phys. {\bf 12}, 87-174 (1940), 
errata: Rev. Mod. Phys. {\bf 13}, 72-73 (1941).

\bibitem{Klyar57} 
A.S. Pokrovskaya-Soboleva and B.N. Klyarfeld,
Sov. Phys. JETP {\bf 5}, 812-818 (1957).

\bibitem{McClure} 
G.W. McClure,
Phys. Rev. {\bf 124}, 969-982 (1961).

\bibitem{Klyar66} 
B.N. Klyarfeld, L.G. Guseva, and A.S. Pokrovskaya-Soboleva,
Sov. Phys. Tech. Phys. {\bf 11}, 520 (1966).

\bibitem{Crowe} R.W. Crowe, J.K. Bragg, and V.G. Thomas,
Phys. Rev. {\bf 96}, 10-14 (1954).

\bibitem{Ward61}
A.L. Ward and E. Jones,
Phys. Rev. {\bf 122}, 376-380 (1961). 

\bibitem{Ward62} 
A.L. Ward, 
J. Appl. Phys. {\bf 33}, 2789-2794 (1962).

\bibitem{Kolobov} 
V.I. Kolobov and A. Fiala,
Phys. Rev. E {\bf 50}, 3018-3032 (1994).

\bibitem{Meek/Craggs} 
J.M. Meek and J.D. Craggs, {\bf Electrical Breakdown of Gases},
(John Wiley and Sons, 1978).

\bibitem{Boeuf} 
J.-P. Boeuf,
J. Appl. Phys. {\bf 63}, 1342-1349 (1988).

\bibitem{FialaPB}
A. Fiala, L.C. Pitchford, and J.-P. Boeuf,
Phys. Rev. E {\bf 49}, 5607-5622 (1994).

\bibitem{pdp} 
Han.S. Uhm, Eun.H. Choi, Guang.S. Cho, 
Appl. Phys. Lett. {\bf 78}, 592 (2001).

\bibitem{Manuel} We thank Dr. Manuel Array\'as for this suggestion.

\bibitem{PhelpsPRE97}
Z.Lj. Petrovic, A.V. Phelps, Phys. Rev. E {\bf 56}, 5920 -- 5931 (1997). 

\bibitem{Auday} 
G. Auday, Ph. Guillot, J. Galy, and H. Brunet,
J. Appl. Phys. {\bf 83}, 5917-5921 (1998).

\end{references}
\end{document}